\documentclass[journal,twocolumn,10pt]{IEEEtran}
\IEEEoverridecommandlockouts 
\usepackage{epsfig,latexsym}
\usepackage{float}
\usepackage{indentfirst}
\usepackage{amsmath}
\usepackage{amssymb}
\usepackage{times}
\usepackage{subfigure}
\usepackage{pifont}
\usepackage{psfrag}
\usepackage{cite}
\usepackage{url}
\usepackage{lastpage}
\usepackage{subfloat}
\usepackage{bm}
\usepackage{color}
\usepackage{fancyhdr}
\usepackage{caption}
\usepackage{balance}
\usepackage{fancyhdr,lastpage}
\usepackage{hyperref}
\usepackage{epstopdf}
\begin{document}
\title{A Novel Transmission Policy for Intelligent Reflecting Surface Assisted Wireless Powered Sensor Networks}
\author{\IEEEauthorblockN{Zheng Chu,
    \emph{Member, IEEE}, 
		Pei Xiao,
  \emph{Senior Member, IEEE}, 
       De Mi,
 \emph{Member, IEEE},
		 Wanming Hao
   \emph{Member, IEEE} \\
    Mohsen Khalily,
    \emph{Senior Member, IEEE}, 
    and 
    Lie-Liang Yang, \emph{Fellow, IEEE}
	}
}
\vspace{-0.2in}
\maketitle 
\thispagestyle{empty}
\begin{abstract}
	This paper proposes a novel transmission policy for an intelligent reflecting surface (IRS) assisted wireless powered sensor network (WPSN). In particular, an energy station (ES) broadcasts energy wirelessly to multiple sensor nodes, which transmit their own information to an access point using the harvested energy. An IRS is deployed to enhance the performance of wireless energy transfer (WET) and wireless information transfer (WIT) by intelligently adjusting phase shifts of each reflecting elements. 
	To achieve its self-sustainability, the IRS needs to collect energy from the ES to support its control circuit operation. Our proposed policy for the considered WPSN is called \emph{IRS assisted harvest-then-transmit time switching} (IRS-HTT-TS), which is able to schedule the transmission time slots by switching between energy collection and energy reflection modes. We study the performance of the proposed transmission policy in terms of the achievable sum throughput, and investigate a joint design of the transmission time slots, the power allocation, as well as the discrete phase shifts of the WET and WIT. This formulates the problem as a mixed-integer non-linear program (MINLP), which is NP-hard and non-convex.
	To deal with this problem, we first relax it to the one with continuous phase shifts, and then propose a two-step approach and decompose the original problem into two sub-problems. 
	We solve the first sub-problem with respect to the phase shifts of the WIT in terms of closed-form expression. Next, we consider two cases to solve the second sub-problem. Specifically, for the special case without the circuit power of each sensor node, the Lagrange dual method and the Karush-Kuhn-Tucker conditions are applied to derive the optimal closed-form transmission time slots, power allocation, and phase shift of the WET. Moreover, we exploit the second sub-problem for the general case with the circuit power of each sensor node, which can be solved via employing a semi-definite programming relaxation. The optimal discrete phase shifts can be obtained by quantizing the continuous values.
	Numerical results demonstrate the effectiveness of the proposed policy and validate the beneficial role of the IRS in comparison to the benchmark schemes. 
\end{abstract}
\IEEEpeerreviewmaketitle
\setlength{\baselineskip}{1\baselineskip}
\newtheorem{definition}{Definition}
\newtheorem{fact}{Fact}
\newtheorem{assumption}{Assumption}
\newtheorem{theorem}{Theorem}
\newtheorem{lemma}{Lemma}
\newtheorem{corollary}{Corollary}
\newtheorem{proposition}{Proposition}
\newtheorem{example}{Example}
\newtheorem{remark}{Remark}
\newtheorem{algorithm}{Algorithm}
\section{Introduction}
Massive connectivity and ultra-high data rate have been two of important pillars in the future generation of communication networks \cite{Dobre_IMM_2019}. 
On the other hand, internet of things (IoT) is typically considered as a significant portion of future wireless networks to guarantee multiple types of connectivities for massive IoT devices in a highly spectral/energy efficient fashion \cite{QQWu_WCM_2017,XChen_JSAC_2020}.  
For a generic IoT system, multiple wireless devices (WDs) (e.g., sensor nodes) establish low-power transmission links to an access point (AP), forming a wireless sensor network (WSN). This has been widely applied to various vertical industries, e.g., event detection of emergency services, structural health monitoring, healthcare diagnosis. \cite{Zheng_WPSN_IoT_2018}. In practice, these WDs are very likely to be powered by the finite-capacity battery, and consequently barely have sufficient energy to support their own computational and communication operations. Traditional ways of battery maintenance and replacement can be often costly and sometime infeasible, especially for massive WDs placed in extreme environments, infrastructures, or human bodies \cite{Rui_Zhang_TWC_2013}. 
Thus, the WDs working with a limited battery lifetime is still a bottleneck for future multiple access wireless networks.

In recent years, radio frequency (RF) wireless energy transfer (WET) has been considered to tackle the energy-constrained issue by novel electromagnetic energy harvesting in a wireless fashion \cite{Varshney_ISIT_2008,Rui_Zhang_TWC_2013,Kaibin_Huang_CM_2015,Brnuo_JSAC_2019}. 
Wireless powered communication network (WPCN) has emerged as one of the promising WET solutions to ensure a stable energy supply. It relies on dedicated energy sources (ESs) to provide wireless charging for the WDs via RF WET, such that these WDs have sufficient power to support their wireless information transfer (WIT) \cite{SZBi_WPC_CM_2015,Rui_Zhang_WPCN_CM_2016}. The WPCN adopts a generic protocol called ``\emph{harvest-then-transmit}'' to schedule the WET and WIT transmission time slots to circumvent the potential interference \cite{Rui_Zhang_WPCN_TWC_2014}. Also, the WPCN can benefit from the extended WDs' operational lifetime via wireless charging and from the reduced maintenance cost. Thus, the WPCN is a viable solution for the low power consumption devices to form a wireless powered sensor network (WPSN) \cite{Zheng_WPSN_IoT_2018}.

On the other hand,
various advanced techniques have been developed to improve the achievable data rate, e.g., relaying, massive multiple-input multiple-output (massive MIMO), ultra-dense networks (UDNs), millimetre wave (mmWave) and Terahertz (THz) communications \cite{ZGao_WC_2015,XChen_CM_2015,SChen_WC_2016}. However, these techniques always incur very high energy consumption and hardware cost, since a large amount of RF chains are used at the transmitter over a high frequency band,  and thus more transmitted data requires more emission of radio waves \cite{RMarcoDi_EURASIP_IRSTUT_2019,Renzo_JSAC_2020}. This has driven the development of a novel and promising paradigm, named \emph{smart and reconfigurable radio environment}. It is a \emph{holographic} wireless mode with an unconventional hardware architecture which features low cost, small size, light weight and low power consumption \cite{CHuang_WCM_2020}. 
\emph{Smart and reconfigurable radio environment} potentially provides a seamless wireless connectivity and the capabilities of transmitting and processing data by recycling the existing radio waves rather than generating new ones, which is a transformative way to convert the traditional wireless environment into a programmable intelligent entity \cite{RMarcoDi_EURASIP_IRSTUT_2019,Renzo_JSAC_2020,CHuang_WCM_2020}.
 As a key enabler of \emph{smart and reconfigurable radio environment}, intelligent reflecting surface (IRS) (also known as reconfigurable intelligent surface (RIS)), has been proved to effectively improve the spectral and energy efficiencies of wireless systems \cite{QWu_CM_2019}. The IRS consists of a large number of reconfigurable reflecting elements, which are managed by a smart controller. These elements possess small-size, low-cost, and low-energy consumption features, and can help strengthen the signal reception by intelligently adjusting the desired signal phase without a dedicated RF processing, en/de-coding, or re-transmission \cite{SHu_TSP_2018}. 
\subsection{State-of-the-Art}
\subsubsection{IRS Assisted Wireless Networks}
Recently, variety of research contributions focus on the IRS-enabled wireless networks \cite{QWu_TWC_2019,CHuang_TWC_EEIRS_2019,ZCHU_WCL_IRS_2019,HSHEN_CL_2019,XGUAN_IRS_AN_2019,CPAN_MCELL_2019,GZhou_TSP_2020,QWu_WCL_2020,QWu_JSAC_2020,CPAN_IRSSWIPTMIMO,WHao_IRS_2020,ibrahim2021exact,zarandi2021delay,Bjornson_CM_20201}.  In \cite{QWu_TWC_2019}, an IRS assisted multiple-input single-output (MISO) downlink system was proposed to jointly optimize the active transmit beamforming of the base station (BS) and passive reflecting beamforming generated by the IRS. The minimization problem of the total transmit power is resolved by an alternating optimization (AO) algorithm to suggest the beneficial role of the IRS in improving power efficiency compared with the traditional schemes. 
The work in \cite{CHuang_TWC_EEIRS_2019} considered a similar system and maximized its energy efficiency such that the power allocation and the phase shifts of the IRS are optimally designed by the AO algorithm, gradient descent search, and sequential fractional programming. 
The integration of IRS with wireless security has been considered in \cite{ZCHU_WCL_IRS_2019,HSHEN_CL_2019,XGUAN_IRS_AN_2019}. The IRS is introduced to reduce the power consumption and enhance the achievable secrecy performance via alternately designing the active secure transmit and passive reflecting beamformers \cite{ZCHU_WCL_IRS_2019,HSHEN_CL_2019}. It was shown in \cite{XGUAN_IRS_AN_2019} that the use of artificial noise (AN) is beneficial to enhance the achievable secrecy rate of an IRS assisted downlink secure system in comparison to its counterparts without IRS or AN. In \cite{CPAN_MCELL_2019}, the IRS is deployed to assist a multi-cell MIMO system, where the edge users can benefit from such a system deployment to mitigate the inter-cell interference. In addition, a block coordinate descent (BCD) algorithm is proposed in \cite{CPAN_MCELL_2019} to solve the formulated weighted sum rate (WSR) maximization, which jointly designs the transmit precoding matrices and phase shifts. The Majorization-Minimization
(MM) algorithm and the complex circle manifold (CCM)
method are adopted to design the locally optimal phase shifts. 
Moreover, an IRS-assisted multigroup multicast system was investigated in \cite{GZhou_TSP_2020}, where a multi-antenna BS broadcasts independent data streams to multiple groups. All single-antenna users in the same group receive the same information, and the inter-group interference is incurred. The second-order cone programming (SOCP) is considered to solve the sum rate maximization of all multicasting groups, and a low complexity algorithm is designed to derive the closed-form solutions of the transmit beamforming and phase shifts. Moreover, the IRS is deployed for not only information reflection to assist the direct link, but also the energy reflection to improve the self-sustainability of wireless networks. The IRS assisted simultaneous wireless information and power transfer (SWIPT) has been investigated in \cite{QWu_WCL_2020,QWu_JSAC_2020}. 
In \cite{QWu_WCL_2020}, by solving the maximization of the weighted harvested energy of energy harvesting users (ERs) subjecting to the individual signal-to-interference-plus-noise ratio (SINR) constraint for information decoding users (IRs), the WET efficiency is enhanced and the rate-energy trade-off is characterized for the IRS enabled SWIPT system. Furthermore, multiple IRSs are deployed to support the information/energy transfer from an access point (AP) to IRs as well as ERs \cite{QWu_JSAC_2020}. The minimization problem of the total transmit power is iteratively solved by an efficient penalty-based method, which demonstrates the effect of IRSs on the energy efficiency enhancement. In \cite{CPAN_IRSSWIPTMIMO}, the maximization problem of the WSR of the IRs was studied for an IRS assisted MIMO SWIPT system, guaranteeing the energy harvesting requirement of the ERs. 
The IRS is integrated in the mmWave and THz communications over  orthogonal frequency division multiple access (OFDMA) in \cite{WHao_IRS_2020}, where the WSR is maximized for the design of the hybrid analog/digital beamforming at the BS and passive beamforming at the IRS. In addition, a more practical case was exploited with imperfect channel state information (CSI), and a robust beamforming and reflecting phase shift matrix are designed to solve the formulated WSR maximization problem. 
An exact coverage analysis of IRS has been investigated with generic Nakagami-m fading channels \cite{ibrahim2021exact}, where a moment generation functions (MGF) based framework was developed to characterize the coverage probability. Joint design for the communication and computing is a promising paradigm \cite{zarandi2021delay}, which fits within the IRS assisted wireless networks to improve offloading efficiency.
In \cite{Bjornson_CM_20201}, RIS, as an option of new physical layer techniques beyond fifth-generation (5G) network, has been overviewed, where three challenges on the controllable wireless environment, better asymptotic array gains, as well as the path loss of the cascaded channel has been unveiled. In addition, two critical questions on use cases and channel estimation with real-time feedback have been unlocked to safeguard the success of RIS. 

\subsubsection{Wireless Powered Communication Network} 
On the other hand, there has been various research endeavours that investigated WPCN \cite{Rui_Zhang_WPCN_TWC_2014,Bruno_TVT_2020,shen2020joint,Slim_WCL_2020,Slim_TCOM_2021,clerckx2021wireless}. In \cite{Rui_Zhang_WPCN_TWC_2014}, the concept of WPCN was proposed, where the dedicated energy supply radiates wireless energy to the energy-constrained users that utilize harvested energy to support information transmissions. Recently, a novel non-linear energy harvesting model was characterized \cite{Bruno_TVT_2020}, where the relation between the direct current (DC) and the received RF signal power was exploited based on its convex property. In \cite{shen2020joint}, a MIMO WET system was investigated to enlarge the output DC power by jointly optimizing the multi-sine waveform and beamforming. It was shown that the proposed scheme offers a higher output DC power than the only beamforming design. In \cite{Slim_WCL_2020}, the impact of the WPCN was studied in the static and mobile coexisting networks, where the total received power and the throughput achieved by the energy harvesting node were characterized. Ambient RF energy harvesting was considered as an enabler of wireless powered IoT networks \cite{Slim_TCOM_2021}, which exploited the packet transmission in grant-free opportunistic uplink WIT charged by the downlink WET. Very recently, new challenges and opportunities for the WET were introduced for future wireless networks to take full advantage of the RF radiations and spectrum in providing cost-effective and real-time power supplies to wireless devices and enable wireless powered applications \cite{clerckx2021wireless}. The benefits of IRS to WET arise from the fact that IRS can help increase the RF power level at the input of the rectenna. 
Thus, the synergy between IRS and WPCN forges a promising solution to improve the energy/information reflection efficiency. Particularly, an IRS was considered in \cite{YShi_IOT_2020} to improve wireless data aggregation in an over-the-air computation (AirComp) system, where 
the minimization problem of the mean-squared-error (MSE) is formulated to jointly design energy and aggregation beamformers at the AP, downlink/uplink phase-shifts at the IRS, and transmit power at the IoT devices, quantifying the AirComp distortion. 

Although the existing works made variety of research contributions in the IRS aided wireless networks, and the IRS aided SWIPT systems, there still exists a major research gap on investigation of the IRS’s benefits in a WPSN. Specifically, utilizing IRS to enhance energy harvesting and data transmission capabilities of the WPSN due to its self-sustainability, has not yet been investigated in the existing works. Furthermore, the IRS can be considered to properly coordinate the energy/information RF signals to offer an coverage enhancement for downlink wireless charging and uplink throughput improvement for the WPSN. Moreover, in the existing works, it is assumed that the IRS does not require any energy for its own circuit operation. However, the IRS controller typically coordinates the channel estimation mode and the data reflection mode \cite{QWu_TWC_2019}, necessitating the sufficient energy for circuit operation. To the best of the authors' knowledge, 
there is few of work having exploited the transmission policy for the IRS aided WPSN, which motivates this paper.

In this paper, a novel transmission policy of a IRS assisted WPSN is proposed, where the IRS helps multiple sensor nodes to enhance their energy harvesting efficiencies and data transmission capabilities, and simultaneously maintain its own circuit operation. 
\emph{The main contributions of this paper are highlighted as follows.}
\begin{enumerate}
	\item We exploit an IRS assisted WPSN system, to be specific, an ES provides WET to multiple sensor nodes, which in turn utilize the collected energy to support their WIT via the time division multiple access (TDMA). Also, the IRS is deployed to improve energy reception at the sensor nodes and information reception at the AP, apart from maintaining its own circuit energy consumption. An \emph{IRS assisted harvest-then-transmit time switching} (IRS-HTT-TS) transmission policy is proposed to efficiently schedule the transmission time slots of the WET and WIT phases. 
	\item To evaluate the performance of the IRS assisted WPSN, we aim to maximize the sum throughput subject to the power constraints of individual sensor node and the IRS, the transmission time constraints, the constraint of the discrete phase shifts. Due to quantized phase shifts, the formulated problem is a mixed-integer non-linear program (MINLP), which is an NP-hard problem. We first relax it into its counterpart with continuous phase shifts which is still a non-convex problem due to multiple coupled variables. 
	\item To deal with the non-convex issue, we propose a two-step approach to decompose the relaxed problem into two sub-problems, which can be solved separately. First, we independently solve the first sub-problem to derive the optimal close-form phase shifts of the WIT. Then, we consider two cases to solve the second sub-problem: a special case and a general case. For the special case without circuit power of each sensor node, we exploit the Lagrange dual method and Karaush-Kuhn-Tucker (KKT) conditions to derive the optimal closed-form transmission time slots, power allocation, and phase shift of the WET. Moreover, we resolve the second sub-problem for the general case with the circuit power of each sensor node, which can be solved by taking into consideration a semi-definite programming (SDP) relaxation. Consequently, the optimal discrete phase shifts can be obtained by quantizing the continuous counterparts. 
\end{enumerate}
The rest of this paper is organized as follows. The system model and the proposed \emph{IRS-HTT-TS} transmission policy are described in Section \ref{section system model}. Section \ref{section:IRSHTTTSPS} investigates the sum throughput maximization for the IRS assisted WPSN. Section \ref{section:Numerical} provides numerical results to evaluate the proposed algorithm. Finally, we conclude this paper in Section \ref{section:Conclusion}.
\subsection{Notations}
We use the upper case boldface letters for matrices and lower case boldface letters for vectors. 
$ \textrm{conj}(\cdot) $, $ (\cdot)^{T} $ and $ (\cdot)^{H} $ denote the conjugate, the transpose and conjugate transpose operations, respectively. $ \textrm{Tr}(\cdot) $ stands for trace of a matrix.
$ \mathbf{A} \succeq  \mathbf{0} $ indicates that $ \mathbf{A} $ is a positive semidefinite matrix. $| \cdot |$ and $ \| \cdot \| $ denote the absolute value and  the Euclidean norm of a vector. 
$ \exp(\cdot) $ and $  \arg(\cdot) $ indicate the exponential function and the phase operation, respectively. $ \mathcal{W}(\cdot) $ denotes the Lambert $ \mathcal{W} $ function. $ \Re \{ \cdot \} $ represents the real part of a complex number. $\textrm{diag}\{ \cdot \}$ denotes a diagonal matrix. $ x \in \mathcal{CN}(0,1) $ denotes that $ x $ follows the complex Gaussian distribution with zero-mean and unit variance.

\section{System Model}\label{section system model}
\begin{figure}[!htbp]
	\centering
	\includegraphics[scale = 0.42]{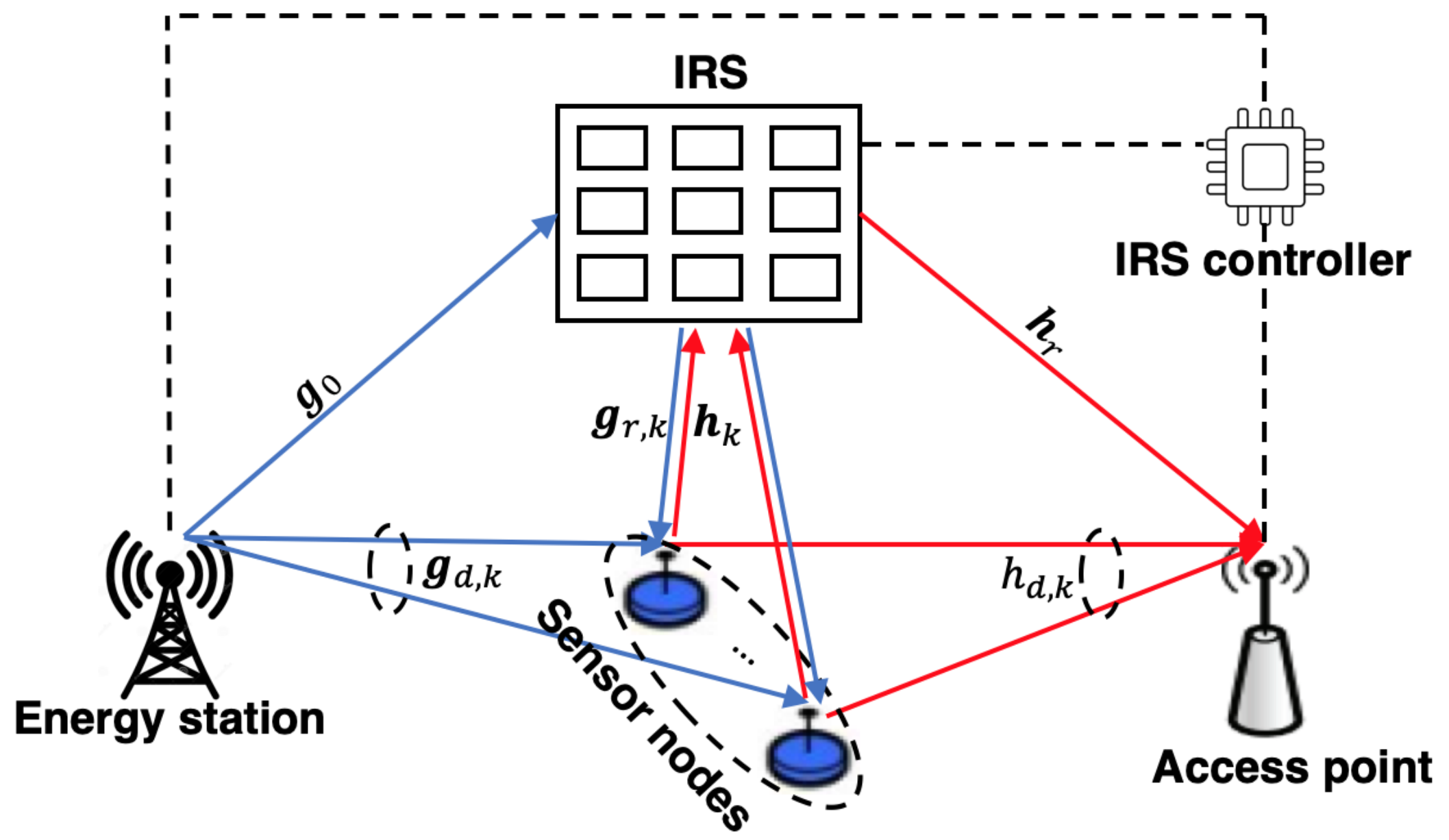}
	\caption{An IRS assisted WPSN.}
	\label{fig:IRS_WPSN}
\end{figure}
\begin{figure}[!htbp]
	\centering
	\includegraphics[scale = 0.45]{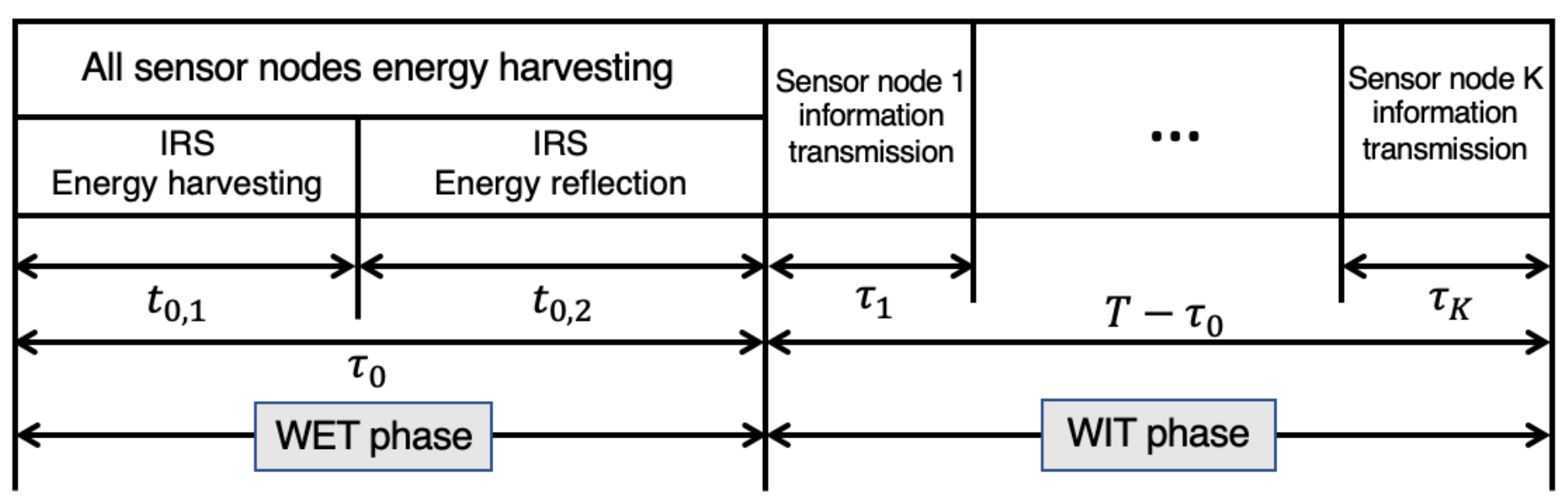}
	\caption{IRS-HTT-TS transmission policy.}
	\label{fig:Time_switching1}
\end{figure}
In this section, we consider an IRS assisted WPSN as shown in Fig. \ref{fig:IRS_WPSN}, which is composed of a single-antenna ES, a single-antenna AP, $ K $ IoT sensor nodes as well as an IRS with $ N_{R} $ passive reflecting elements.
Specifically, the ES, connected with a stable microgrid for energy supply, provides wireless charging service to the sensor nodes, that can utilize the harvested energy to transmit their own message to the AP.\footnote{Although the work in \cite{BLyu_2020} considered a similar system model, it integrated the information receiving station (i.e., AP) and the ES into one hybrid AP (HAP), which results in a lower hardware complexity but induces the doubly near-far issue \cite{QQWU_EEWPCN_TWC_2016}. In \cite{BLyu_2020}, the authors employed the \emph{well-known} methods, such as one-dimensional line search and SDP relaxation, to numerically design the second sub-slot of the WET phase as well as jointly optimize the transmission time scheduling of the WIT phase and the phase shifts of the WET phase, respectively. In our work, a different modelling and a novel scheme are proposed, and in the following sections, we will highlight the advantages of our work by showing how it can effectively reduce the computational complexity.}
 Coordinated by an IRS controller, the IRS supports passive energy/information beamforming to improve the energy/information transmission efficiency. The IRS requires a portion of energy consumption to support its own circuit operation.
To schedule the transmission time slots, a whole operation time block is set to $ T $, which can be split into two time periods, WET phase and WIT phase. During the WET phase, the IRS  collects a portion of energy to support its own circuit operation and reflects the rest to improve energy reception of the sensor nodes.
During the WIT phase, all sensor nodes utilize the harvested energy to deliver their own information to the AP with the IRS's help and in TDMA mode.\footnote{By setting the corresponding biasing voltages of the IRS varactors or PIN diodes or changing the value of resistors in each element, we can achieve the controllable reflection coefficients, i.e., phase shift and refection amplitude, in every time slot of TDMA \cite{QWu_CM_2019}. }
We denote $ \tau_{0} $ as the time duration for the WET phase, and $ \tau_{k},~\forall k \in [1,K] $ as the time duration for the $ k $-th sensor nodes during the WIT phase, all of which satisfies  $ \sum_{k=0}^{K} \tau_{k} = T $.
To schedule the transmission time slots of WET phase, 
we adopt a time switching mode,
where the first sub-slot is considered for the energy harvesting mode to support the circuit energy consumption of the IRS, and the second sub-slot is for the energy reflection mode to improve energy harvesting at the sensor nodes. 
This proposed novel transmission policy, named, IRS-HTT-TS is illustrated in Fig. \ref{fig:Time_switching1}.  
In what follows, we focus on developing the suitable optimization strategies for the proposed transmission policy, and we are interested in the upper bounds of the sum throughput performance of different strategies for the considered system model. Therefore, perfect channel state information (CSI) is assumed in this work. Note that several channel estimation techniques have been investigated for the WPSN in the literature for obtaining the CSI of the direct link between ES and devices as well as that between devices and AP \cite{RZhang_WPCN_TWC_2014,QQWU_EEWPCN_TWC_2016}. 
 One can also consider the passive pilots used for the channel estimation of the cascaded CSI. In particular, the reflecting elements of the IRS passively reflect the pilot sequences transmitted from the IoT devices to the AP/ES to the IoT devices such that the CSI associated with the IRS can be obtained \cite{LLiu_TWC2020}. 
 On the other hand, the imperfect CSI cases have been investigated in \cite{CPAN_Robust1_2020,CPAN_Robust2_2020,CPAN_Robust3_2020}, showing that the uncertainty of CSI or channel estimation errors can degrade system throughput. These imperfections can be tackled by using robust resource allocations for active transmit and passive reflecting beamformers \cite{CPAN_Robust1_2020,CPAN_Robust2_2020,CPAN_Robust3_2020}. To be specific, these robust schemes take into account the imperfect cascaded BS-IRS-user channels, which typically follow bounded and statistical CSI uncertainty, and the worst-case as well as outage probability robust beamforming designs are characterized.
In addition, the channel coefficients between the ES and the $ k $-th IoT device, the ES and the IRS, the IRS and the $ k $-th IoT device, the $ k $-th IoT device and the AP, the $ k $-th IoT device and the IRS, as well as the IRS and the AP are denoted by $ g_{d,k} \in \mathbb{C}^{1 \times 1} $, $ \mathbf{g}_{0} \in \mathbb{C}^{1 \times N_{R}} $, $ \mathbf{g}_{r,k} \in \mathbb{C}^{N_{R} \times 1} $, $ h_{d,k} \in \mathbb{C}^{1 \times 1} $,  $ \mathbf{h}_{k} \in \mathbb{C}^{1 \times N_{R}} $, and $ \mathbf{h}_{r} \in \mathbb{C}^{N_{R} \times 1} $, respectively. 
We assume that all channels are quasi-static flat fading, and stay constant during one time block but may vary from one to another. Also, the perfect CSI for all transmission links are available to evaluate the upper bound for the system throughput.


\subsection{Phase Shift Description}
For the IRS, we denote $ \bm{\Theta}_{k} = \textrm{diag} \left[\beta_{k,1}\exp(j\alpha_{k,1}), ..., 
 \beta_{k,N}\exp(j\alpha_{k,N_{R}})\right], \forall k \in [0,K], n \in [1,N_{R}] $ as the diagonal phase shift matrix for different time durations, where $ \alpha_{k,n} \in [0, 2 \pi) $ and  $ \beta_{k,n} \in [0, 1] $ are the phase shift and reflection coefficient of the associated IRS reflecting elements, respectively. The phase shift of each reflecting element satisfies $ | \exp(j\alpha_{k,n}) | = 1 $. When $ k = 0 $, $ \bm{\Theta}_{0} $ is also known as the energy reflection phase shift matrix, whereas $ \bm{\Theta}_{k} $ is called the information reflection phase matrix when $ k \in [1,K] $. 
 Since the IRS is deployed to participate in the WET and WIT, aiming to maximize the energy and information reflection, the reflecting amplitude of each elements on the IRS should be set to a value of $ \beta_{k,n} = 1, k \in [0,K], n \in [1,N_{R}] $ to achieve the maximum reflection gain.
 In practice, the phase shift of each element is designed to be finite number of discrete values, which can uniformly quantize the set of interval $ [0,2\pi) $ \cite{QWu_Discrete_TCOM2020}, i.e., 
\begin{align}\label{eq:Dis_phaseshift_constraint}
\!\! \mathcal{S}_{\bm{\Theta}} &~\!=\! \left\{ \theta_{k,n} \!=\! \exp(j \alpha_{k,n}), \alpha_{k,n} \! \in \! \left\{  0, \frac{2 \pi}{ L }, ..., \frac{2 \pi \left( L \!-\! 1 \right)}{ L } \right\} \right\}, \nonumber\\ 
&~ \forall k \in [0,K],~\forall n \in [1,N_{R}],
\end{align}
where $ \mathcal{S}_{\bm{\Theta}} $ denotes the set of finite phase shifts, $ L = 2^{B} $ is the total number of phase shift levels, and $ B $ denotes the number of bits used to indicate the number of phase resolutions.

 
\subsection{Signal Processing of IRS-HTT-TS}\label{section:signal_description}
Here, we describe the IRS-HTT-TS transmission policy from the signal processing perspective. 
\subsubsection{Wireless Energy Transfer of IRS-HTT-TS Transmission Policy}
As shown in the left part of Fig. \ref{fig:Time_switching1}, the whole WET time slot is denoted by $ \tau_{0} $, which can be split into two sub-slots, $ t_{0,1} $ and $ t_{0,2} $, and $ \sum_{i = 1}^{2} t_{0,i} \leq \tau_{0} $. The sensor nodes collect wireless energy from the ES over the whole WET time slot. Also, the IRS harvests energy from the ES during the first sub-slot $ t_{0,1} $ for its own circuit operation, and then it reflects energy from the ES to the sensor nodes during the second sub-slot $ t_{0,2} $ to strengthen the energy reception at the sensor nodes. 
Thus, the received signals at the IRS during $ t_{0,1} $, as well as at the $ k $-th sensor node during $ t_{0,1} $ and $ t_{0,2} $, respectively, are given by 
\begin{align}
\mathbf{y}_{r}^{\textrm{TS}} & \!=\! \sqrt{P_{0}} \mathbf{g}_{0}^{H} x_{0} \!+\! \mathbf{n}_{\textrm{IRS}},   \nonumber \\
y_{k,i}^{\textrm{TS}} & \!=\! \left\{
\begin{array}{lcl}
\!\!\!\! \sqrt{P_{0}} g_{d,k} x_{0} \!+\! n_{k},      &  \!\!\!\!\!\!\!\!\!    &\!\! \textrm{During}~ t_{0,1}, \\
\!\!\!\! \sqrt{P_{0}} \left( \mathbf{g}_{0} \bm{\Theta}_{0} \mathbf{g}_{r,k} \!+\! g_{d,k} \right) x_{0} \!+\! n_{k},   &    \!\!\!\!\!\!\!\!\! & \!\! \textrm{During} ~t_{0,2},
\end{array} \right. \!\!\!\!, \forall k \! \in \! [1,K],
\nonumber
\end{align}
where $ x_{0} $ denotes the energy signal with unit power, i.e.,  $ x_{0} \in \mathcal{CN}(0,1) $, at the ES;
$ P_{0} $ denotes the transmit power at the ES; $ n_{k} $ and $ \mathbf{n}_{\textrm{IRS}} $ are the additive whit Gaussian noises at the $ k $-th sensor node and the IRS, respectively.\footnote{Note that the noise power of the WET phase is typically ignored due to its negligible impact on energy harvesting. It is assumed based on the fact that the transmit power at the ES $ P_{0} $ is sufficiently large \cite{Rui_Zhang_WPCN_TWC_2014}.}
Accordingly, the energy collected by the IRS during $ t_{0,1} $ and by the $ k $-th sensor node during $ t_{0,1} $ and $ t_{0,2} $, 
respectively, can be represented by
$
E_{\textrm{TS,IRS}} \!=\! t_{0,1} \eta P_{0} \left \| \mathbf{g}_{0} \right \|^{2} $, and $
E_{\textrm{TS},k} \!=\! t_{0,1} \eta P_{0} \left | g_{d,k} \right |^{2} \! +\! t_{0,2} \eta P_{0} \left | \mathbf{g}_{0} \bm{\Theta}_{0} \mathbf{g}_{r,k} \!+\! g_{d,k} \right |^{2}, \forall k \in [1,K], $
where $ \eta $ denotes the energy harvesting efficiency. 

\subsubsection{Wireless Information Transfer of IRS-HTT-TS}
During the WIT phase, all sensor nodes use the harvested energy to transmit their own message to the AP under the TDMA protocol, with the assistance of the IRS.
Let $ P_{k} $, $ P_{c,k} $ and $ P_{c,\textrm{IRS}} $ denote the transmit power, the circuit power consumption at the $ k $-th sensor node, and the circuit power consumption at the IRS, respectively. 
In this paper, we consider the energy harvesting at the IRS to support its own circuit operation and at the IoT sensors to support their individual information transmission as well as circuit operation. For these purpose, two different relations are considered between the harvested energy and the required energy. To be specific, for the IRS, it follows the fact that the total energy consumption to support the IRS circuit operation during the second sub-slot of the WET phase and the whole WIT phase cannot exceed the harvested energy during the first sub-slot of the WET phase. 
For the IoT sensors, the harvested energy is consumed by each IoT sensor not only for supporting its information transmission, but also for supporting its circuit operation. Thus, the energy constraint on the $ k $-th sensor is  
\begin{align}\label{eq:HTT_TS_sensorpower}
\tau_{k} \left( P_{k} + P_{c,k} \right) \leq E_{\textrm{TS},k},
~\forall k \in [1,K].
\end{align}   
To achieve self-sustainability, it should be guaranteed that the total energy consumption of the IRS's circuit operation does not exceed its harvested energy. Hence, the corresponding constraint can be written as 
\begin{align}\label{eq:HTT_TS_IRSpower}
N_{R} P_{c,\textrm{IRS}} \left( t_{0,2} + \sum_{k=1}^{K} \tau_{k} \right) \leq E_{\textrm{TS,IRS}}.
\end{align}
In addition, the received signal from the $ k $-th sensor node at the AP is given by 
\begin{align}
y_{\textrm{AP},k} = \sqrt{P_{k}} \left( \mathbf{h}_{k} \bm{\Theta}_{k} \mathbf{h}_{r} + h_{d,k} \right) x_{k} + n_{\textrm{AP}},
\end{align}
where $ x_{k} \sim \mathcal{CN} (0,1) $ denotes the information signal sent by the $ k $-th sensor node; $ n_{\textrm{AP}} \sim \mathcal{CN} \left(0,\sigma^{2}\right) $ denotes the AWGN at the AP. 
Thus, the achievable throughput of the $ k $-th sensor node at the AP can be calculated as 
\begin{align}\label{eq:Rk}
R_{k} 
 = \tau_{k} \log \left( 1 + \frac{ P_{k} \left | \mathbf{h}_{k} \bm{\Theta}_{k} \mathbf{h}_{r} + h_{d,k} \right |^{2} }{ \sigma^{2} }  \right)
\end{align}
\subsection{Problem formulation}  
To evaluate the IRS-HTT-TS transmission policy, we aim to maximize the achievable sum throughput, with a joint design of the phase shifts of the WET and WIT phases, the transmission time scheduling, as well as the transmit power allocation of all sensor nodes. The formulated problem is given as 
 \begin{subequations}\label{eq:Ori_problem_HTT_TS0}
		\begin{align}
		&\!\!\!\!\!\!\!\!\!\!\! \!\! \max_{\bm{\Omega}_{\textrm{IRS-HTT-TS}}} ~ \sum_{k=1}^{K} \tau_{k} \log \left( 1 + \frac{ P_{k} \left | \mathbf{h}_{k} \bm{\Theta}_{k} \mathbf{h}_{r} + h_{d,k} \right |^{2} }{ \sigma^{2} }  \right),  \label{eq:Ori_problem_HTT_TS_obj}\\ 
      s.t. &~ 
	 \eqref{eq:HTT_TS_sensorpower},
		~ 
		\eqref{eq:HTT_TS_IRSpower}\\
		&~ \theta_{k,n} \in \mathcal{S}_{\bm{\Theta}},~\forall k \in [0,K],~\forall n \in [1,N_{R}], \label{eq:HTT_phase} \\
		&~ \sum_{i = 1}^{2} t_{0,i} \leq \tau_{0}, \label{eq:HTT_TS_time}\\ 
		&~ \sum_{k=0}^{K} \tau_{k} \leq T, \label{eq:HTT_TS_timeallocation0}\\
		&\!\!\!\!\!\!\!\!\!\!\!  t_{0,i} \geq 0, \forall i \in [1,2],\tau_{k} \geq 0, \forall k \in [0,K],P_{k} \geq 0, \forall k \in [1,K], \label{eq:Variables1} 
		\end{align}
	\end{subequations} 
 where $ \bm{\Omega}_{\textrm{IRS-HTT-TS}} \!=\!\! \left[\!  \left\{ \bm{\Theta}_{k} \right\}_{0}^{K}\!, \left\{ t_{0,i} \right\}_{1}^{2}\!, \left\{ \tau_{k} \right\}_{0}^{K}\!, \left\{ P_{k} \right\}_{1}^{K}  \right] $ assembles all variables of problem \eqref{eq:Ori_problem_HTT_TS0}.
In problem \eqref{eq:Ori_problem_HTT_TS0}, \eqref{eq:HTT_TS_sensorpower} denotes the transmit power constraint of the $ k $-th sensor node for the IRS-HTT-TS policy; \eqref{eq:HTT_TS_IRSpower} is the IRS power constraints for these policies; constraint \eqref{eq:HTT_phase} denotes the discrete phase shift constraint;
\eqref{eq:HTT_TS_time} is the constraint of time scheduling during the WET; Constraint \eqref{eq:HTT_TS_timeallocation0} 
denotes the total time constraint; 
\eqref{eq:Variables1} lists individual transmission time and power constraints.
In our system model, the AP is typically equipped with multiple antennas, which employs the receive beamforming to decode the information received from different IoT sensors, instead of using TDMA.
The associated problem formulation is given as \eqref{eq:Rk1} on the top of next page,
\begin{figure*}
\begin{align}\label{eq:Rk1}
	\max_{ \bm{\Omega}_{\textrm{IRS-HTT-TS}} } &~ \sum_{k=1}^{K} \tau_{1} \log \left( 1 + \frac{ P_{k} \left | \mathbf{w}_{k} \left( \mathbf{H}_{r} \bm{\Theta}_{1} \mathbf{h}_{k} + \mathbf{h}_{d,k} \right) \right |^{2} }{ \mathbf{w}_{k} \left[ \sum_{j \neq k} P_{j}    \left( \mathbf{H}_{r} \bm{\Theta}_{1} \mathbf{h}_{j} + \mathbf{h}_{d,j} \right) \left( \mathbf{H}_{r} \bm{\Theta}_{1} \mathbf{h}_{j} + \mathbf{h}_{d,j} \right)^{H}   + \sigma^{2}\mathbf{I} \right] \mathbf{w}_{k}^{H} }   \right), \nonumber\\ 
	s.t.&~ \sum_{i = 1}^{2} t_{0,i} \leq \tau_{0}, ~ \sum_{k=0}^{1} \tau_{k} \leq T, 
	~ t_{0,i} \geq 0, ~\forall i \in [1,2],~\tau_{k} \geq 0, ~\forall k \in [0,1],~P_{k} \geq 0, ~\forall k \in [1,K],  \nonumber\\ 
	&~ \left | \exp \left( j \alpha_{k,n} \right) \right |  \!= \! 1, \alpha_{k,n}  \! \in \! [0, 2\pi),  \forall k  \! \in \! [0, K], \forall n  \! \in \! [1, N_{R}], \nonumber\\ 
	&~ \tau_{1} \left( P_{k} + P_{c,k} \right) \leq t_{0,1} \eta P_{0} \left | g_{d,k} \right |^{2} \! +\! t_{0,2} \eta P_{0} \left | \mathbf{g}_{0} \bm{\Theta}_{0} \mathbf{g}_{r,k} \!+\! g_{d,k} \right |^{2},
	~\forall k \in [1,K], \nonumber\\ 
	&~ N_{R} P_{c,\textrm{IRS}} \left( t_{0,2} +  \tau_{1} \right) \leq t_{0,1} \eta P_{0} \left \| \mathbf{g}_{0} \right \|^{2},  
\end{align} 
\end{figure*}
where $ \bm{\Omega}_{\textrm{IRS-HTT-TS}} \!=\!\! \left[\!  \left\{ \bm{\Theta}_{0}, \bm{\Theta}_{1} \right\} \!, \left\{ t_{0,i} \right\}_{1}^{2}\!, \left\{ \tau_{0},\tau_{1} \right\} \!, \left\{ P_{k} \right\}_{1}^{K}  \right] $ assembles all the variables of problem \eqref{eq:Rk1},
$ \mathbf{H}_{r} \in \mathbb{C}^{M \times N_{R}} $, $ \mathbf{h}_{k} \in \mathbb{C}^{N_{R} \times 1} $, and $ \mathbf{h}_{d,k} \in \mathbb{C}^{N_{R} \times 1} $. Note that problem \eqref{eq:Rk1} considers that the AP is equipped with $ M $ antennas such that TDMA is not necessary due to the receive beamformer $ \mathbf{w}_{k},~\forall k \in [1,K] $ for the WIT phase. This formulated problem includes multiple coupled received beamformer $ \mathbf{w}_{k} $ and phase shifts $ \bm{\Theta}_{i},~\forall i \in [0,1] $, which is not convex and cannot be solved directly. 
A possible solution is to employ the AO algorithm to alternately design $ \mathbf{w}_{k} $ and $ \bm{\Theta}_{i} $ \cite{QWu_TWC_2019}. Also, it is not possible to derive the optimal transmission time scheduling and phase shifts of the WET phase in closed-form. This because to achieve these, a more complicated derivation and reformulation have to be considered. However, for this scenario, it deserves another investigation, which is left for our future work. 

\section{Achievable Sum Throughput Maximization for IRS-HTT-TS}\label{section:IRSHTTTSPS}
Problem \eqref{eq:Ori_problem_HTT_TS0} is not jointly convex due to multiple coupled variables in its objective function \eqref{eq:Ori_problem_HTT_TS_obj} and constraint \eqref{eq:HTT_TS_sensorpower}. Also, the phase shifts $ \alpha_{k,n},~\forall k \in[0,K],~\forall n \in [1,N_{R}] $ in constraint \eqref{eq:Dis_phaseshift_constraint} imply that \eqref{eq:Ori_problem_HTT_TS0} is a mixed-integer non-linear program (MINLP), which is typically NP-hard and cannot be solved directly. To deal with this issue, we first relax each discrete phase shift $ \alpha_{k,n} $ to its continuous counterpart, i.e., $ \left | \exp \left( j \alpha_{k,n} \right) \right | = 1,~ \alpha_{k,n} \in [0, 2\pi), ~ \forall k \in [0,K],~\forall n \in [1,N_{R}] $. 
Thus, problem \eqref{eq:Ori_problem_HTT_TS0} is relaxed as 
\begin{subequations}\label{eq:Ori_problem_HTT_TS}
	\begin{align}
	\max_{\bm{\Omega}_{\textrm{IRS-HTT-TS}}} &~  \sum_{k=1}^{K} \tau_{k} \log \left( 1 + \frac{ P_{k} \left | \mathbf{h}_{k} \bm{\Theta}_{k} \mathbf{h}_{r} + h_{d,k} \right |^{2} }{ \sigma^{2} }  \right),  \label{eq:Ori_problem_HTT_TS_obj1}\\ 
	s.t. &~\eqref{eq:HTT_TS_sensorpower},~ \eqref{eq:HTT_TS_IRSpower},~ \eqref{eq:HTT_TS_time},~\eqref{eq:HTT_TS_timeallocation0},~\eqref{eq:Variables1},\\
	& \! \! \! \! \! \! \! \! \! \! \! \! \! \! \! \! \! \! \! \! \! \! \!\!\!\! \left | \exp \left( j \alpha_{k,n} \right) \right |  \!= \! 1, \alpha_{k,n}  \! \in \! [0, 2\pi),  \forall k  \! \in \! [0, K], \forall n  \! \in \! [1, N_{R}]. \label{eq:phase}
	\end{align}
\end{subequations}
Problem \eqref{eq:Ori_problem_HTT_TS} is still non-convex with the multiple coupled variables. 
To circumvent this non-convexity, in the following, we will propose a two-step approach to solve it, where problem \eqref{eq:Ori_problem_HTT_TS} is divided into two sub-problems, which can be separately solved.
\vspace{-1em}
\subsection{Optimal Phase Shifts of WIT}\label{section:Phase_WIT}
In this subsection, we solve the first sub-problem with respect to the phase shifts of the WIT phase, i.e., $ \bm{\Theta}_{k},~\forall k [1,K] $. To proceed, the following \emph{theorem} is in order.
\begin{theorem}\label{lemma:EQphaseshifts}
	The optimal phase shifts of the WIT phase, i.e., $ \bm{\Theta}_{k},~\forall k \in [1,K] $ can be derived as 
	\begin{align}
		\bm{\Theta}_{k}^{*} \!=\! \textrm{diag}\left( \theta_{k,1}^{*}, ..., \theta_{k,N_{R}}^{*}  \right)\!, \forall k \in [1,K],\forall n \in [1,N_{R}],
	\end{align}
where $ \theta_{k,n}^{*} \!=\! \exp \left( j \alpha_{k,n}^{*} \right) $, $ \alpha_{k,n}^{*} \!=\! \arg \left( h_{d,k} \right) \!-\! \arg \left( \mathbf{b}_{k} [n] \right) $, and $ \mathbf{b}_{k} = \textrm{diag}\left( \mathbf{h}_{k} \right) \mathbf{h}_{r} $. 
%
\end{theorem}
\begin{IEEEproof}
	See Appendix \ref{appendix:EQphaseshifts}.
	\end{IEEEproof}
Consequently, the following \emph{lemma} is presented to clarify the impact of the optimal phase shifts, i.e., $ \bm{\Theta}_{k} $ on the signal reception enhancement at the AP. 
\begin{lemma}\label{proposition:Alignment}
	The optimal phase shifts of the WIT phase provide an alignment of the reflecting link between the sensor nodes and the AP via the IRS with the direct link between them, leading to the following relation
	\begin{align}\label{eq:Relation_direct_reflection}
	\mathbf{h}_{k} \bm{\Theta}_{k}^{*} \mathbf{h}_{r} = \omega_{k} h_{d,k},~\forall k \in [1,K],
	\end{align}
where $ \omega_{k} $ is positive scalar. 
\end{lemma}
\begin{IEEEproof}
	See Appendix \ref{appendix:Alignment}.
	\end{IEEEproof}
By exploiting \emph{Lemma} \ref{proposition:Alignment}, with the help of the IRS during the WIT phase, the signal reception of the $ k $-th sensor node at the AP can be strengthened at most $ (1 + \omega_{k})^{2} $ times in comparison to the case without IRS \cite{BLyu_2020}. Actually, $ \omega_{k} $, $ \forall k \in [1,K] $ is proportionate to the reflecting elements of the IRS. Hence, more significant enhancement is introduced in terms of the sum throughput performance with a larger number of reflecting elements. Moreover, the IRS deployment as well as the path loss model of the reflecting link also significantly affect the signal reception at the AP \cite{Bjornson_WCL_2020}.

After obtaining the optimal phase shifts of the WIT phase, we further solve problem \eqref{eq:Ori_problem_HTT_TS} to optimally design the phase shift of the WET phase, the transmission time allocation, the transmit power of the sensor nodes. Denoting $ c_{k} = \left | \mathbf{h}_{k} \bm{\Theta}_{k}^{*} \mathbf{h}_{r} + h_{d,k} \right |^{2},
$ we have 
	\begin{subequations}\label{eq:Ori_problem_HTT_TS1}
		\begin{align}
		\max_{\bm{\Omega}_{\textrm{IRS-HTT-TS}}} &~  \sum_{k=1}^{K} \tau_{k} \log \left( 1 + \frac{ P_{k} c_{k} }{ \sigma^{2} }  \right),  \\ 
		s.t. &~ \eqref{eq:HTT_TS_sensorpower},~\eqref{eq:HTT_TS_IRSpower},~\eqref{eq:HTT_TS_time},~\eqref{eq:HTT_TS_timeallocation0}, \\
		&\!\! \left | \exp \left( j \alpha_{0,n} \right) \right | \!=\! 1, \alpha_{0,n} \! \in \! [0, 2\pi), \forall n \! \in \! [1,N_{R}], \label{eq:phase1} \\
		&\!\!  t_{0,i} \geq 0, ~\forall i \in [1,2],~\tau_{k} \geq 0, ~\forall k \in [0,K],\nonumber\\&~~~~~~~~P_{k} \geq 0, ~\forall k \in [1,K], \label{eq:HTT_TS_variables1}\\
		&\!\!\! \bm{\Omega}_{\textrm{IRS-HTT-TS}} = \left[ \bm{\Theta}_{0}, \left\{ t_{0,i} \right\}_{1}^{2}, \left\{ \tau_{k} \right\}_{0}^{K}, \left\{ P_{k} \right\}_{1}^{K}   \right]. \label{eq:HTT_TS_variables2}
		\end{align}
	\end{subequations}
\subsection{Optimal Solution to Problem \eqref{eq:Ori_problem_HTT_TS1}}\label{section:Optimal_IRS_HTT_TS}
In this subsection, we analyze the throughput performance of the proposed IRS-HTT-TS transmission policy via solving problem \eqref{eq:Ori_problem_HTT_TS1}. To proceed, we begin with the special case (i.e., $ P_{c,k} = 0 $) and then the general case (i.e., $ P_{c,k} \neq  0 $). 
\subsubsection{Special Case ($ P_{c,k} = 0 $)}\label{section:Sepecial_TS}
Here, we consider the special case when the circuit power consumption of the $ k $-th sensor node equals to zero, i.e., $ P_{c,k} = 0 $.\footnote{This case help us to obtain the upper bound of the achievable sum throughput for the proposed IRS-HTT-TS policy. A more general and practical case ($ P_{c,k} \neq 0 $) will be introduced in Section \ref{section:TS_general} that each sensor node collects energy to support its circuit operation during the WET.} 
Thus, problem \eqref{eq:Ori_problem_HTT_TS1} is simplified to
	\begin{subequations}\label{eq:Ori_problem_HTT_TS2}
	\begin{align}
	\max_{\bm{\Omega}_{\textrm{IRS-HTT-TS}}} &~  \sum_{k=1}^{K} \tau_{k} \log \left( 1 + \frac{ P_{k} c_{k} }{ \sigma^{2} }  \right), \label{eq:Ori_problem_HTT_TS2_obj} \\ 
	s.t. &~	 \tau_{k}  P_{k}   \leq E_{\textrm{TS},k},  \label{eq:HTT_TS_specialsensorpower} \\
	&~\eqref{eq:HTT_TS_IRSpower},~\eqref{eq:HTT_TS_time},~\eqref{eq:HTT_TS_timeallocation0},~\eqref{eq:phase1},~\eqref{eq:HTT_TS_variables1},~\eqref{eq:HTT_TS_variables2}.
	\end{align}
\end{subequations}
Problem \eqref{eq:Ori_problem_HTT_TS2} is still non-convex and intractable. To solve it, we first consider the following \emph{lemma} to handle the transmit power of the sensor nodes as well the sub-slots of the WET phase. 
\begin{lemma}\label{lemma:Handle1}
	The optimal transmit power of the $ k $-th sensor node $ P_{k}^{*},~\forall k \in [1,K] $ and the optimal second sub-slot of the WET phase $ t_{0,2}^{*} $ should satisfy the following equalities: 
	\begin{subequations}
\begin{align}
P_{k}^{*} &~ = \frac{ t_{0,1}^{*} \eta P_{0} | g_{d,k} |^{2} + t_{0,2}^{*} \eta P_{0} | \mathbf{g}_{0} \bm{\Theta}_{0}^{*} \mathbf{g}_{r,k} + g_{d,k} |^{2} }{ \tau_{k}^{*} }, \label{eq:Optimal_Pk0}\\ 
t_{0,2}^{*} &~=T - \sum_{k=1}^{K} \tau_{k}^{*}  - t_{0,1}^{*}, \label{eq:Optimal_t020}
\end{align}
\end{subequations}
where $ t_{0,i}^{*}, ~\forall i \in [1,2] $, $ \tau_{k}, ~\forall k \in [1,K] $, and $ \bm{\Theta}_{0}^{*} $ denote the optimal solutions of $ t_{0,i} $, $ \tau_{k} $, and $ \bm{\Theta}_{0} $, respectively. In addition, the optimal first sub-slots of the WET phase $ t_{0,1}^{*} $ can be derived in closed-form, as 
\begin{align}\label{eq:Optimal_t01}
t_{0,1}^{*} = \frac{N_{R} P_{\textrm{c,IRS}} T }{ N_{R} P_{\textrm{c,IRS}} + \eta P_{0} \| \mathbf{g}_{0} \|^{2} }
\end{align}
\end{lemma}
\begin{IEEEproof}
	See Appendix \ref{appendix:Handle1}.
	\end{IEEEproof}
By applying \emph{Lemma} \ref{lemma:Handle1}, and defining $ \tilde{c}_{k} = \frac{ c_{k} \eta P_{0} }{ \sigma^{2} } $, $ \bar{c}_{k} = t_{0,1}^{*} | g_{d,k} |^{2} $ as well as $ s_{k} = | \mathbf{g}_{0} \bm{\Theta}_{0} \mathbf{g}_{r,k} + g_{d,k} |^{2} $, problem \eqref{eq:Ori_problem_HTT_TS2} can be further reformulated as
\begin{subequations}\label{eq:Ori_problem_HTT_TS_special2}
\begin{align}
\max_{ \bm{\Theta}_{0}, t_{0,2}, \left\{ \tau_{k} \right\}_{1}^{K}  } &~  \sum_{k=1}^{K} \tau_{k} \log \left( 1 + \frac{ \tilde{c}_{k}  \left( \bar{c}_{k} + t_{0,2} s_{k} \right) }{ \tau_{k}  } \right),~
s.t.~\eqref{eq:phase1}, \nonumber\\ &~ t_{0,2} \leq T - \sum_{k=1}^{K} \tau_{k} - t_{0,1}^{*}, \label{eq:Ori_problem_HTT_TS_special2_timeconstraint} \\ 
&~ t_{0,2} \geq 0,~\tau_{k} \geq 0, ~\forall k \in [1,K],
 \label{eq:Ori_problem_HTT_TS_special2_timeconstraint1} 
\end{align}
\end{subequations}
Problem \eqref{eq:Ori_problem_HTT_TS_special2} can be relaxed into a semidefinite programming (SDP), and solved directly by employing interior-point methods \cite{boyd_B04}. However, the SDP relaxation may incur a higher rank of the SDP based solution such that the Gaussian randomization technique should be adopted to construct an approximated feasible solution. This may introduce a high computational complexity and is time-consuming. It is thus imperative to develop a low complexity scheme which derives an optimal closed-form solution of the transmission time (i.e., $ t_{0,2} $, and $ \tau_{k},~\forall k \in [1,K] $), as well as the phase shift of the WET phase (i.e., $ \bm{\Theta}_{0} $).
In order to perform the low complexity scheme, we first derive the transmission time of the WIT phase, i.e., $ \tau_{k},~\forall k \in [1,K] $, for which the following \emph{theorem} is required, 
\begin{theorem}\label{theorem:TS_tauk}
	The optimal transmission time of the WIT phase, i.e., $ \tau_{k},~\forall k \in [1,K] $, is derived in closed-form, as 
	\begin{align}\label{eq:tauk1}
	\tau_{k}^{*} = \frac{ \left( T - t_{0,1}^{*} - t_{0,2} \right) \tilde{c}_{k}  \left( \bar{c}_{k} + t_{0,2} s_{k} \right)  }{ \sum_{k=1}^{K} \tilde{c}_{k}  \left( \bar{c}_{k} + t_{0,2} s_{k} \right) }
	\end{align}
\end{theorem}
\begin{IEEEproof}
	See Appendix \ref{appendix:TS_tauk}.
	\end{IEEEproof}
We further substitute \eqref{eq:tauk1} in \emph{Theorem} \ref{theorem:TS_tauk} into \eqref{eq:Ori_problem_HTT_TS_special2}, and define $ \tilde{T} = T - t_{0,1}^{*} $, $ c = \sum_{k=1}^{K} \tilde{c}_{k} \bar{c}_{k} $, as well as $ s = \sum_{k = 1}^{K} \tilde{c}_{k} s_{k} $, then \eqref{eq:Ori_problem_HTT_TS_special2} is equivalent to
\begin{subequations}\label{eq:Ori_problem_HTT_TS_special8} 
\begin{align}
\max_{ t_{0,2}, \bm{\Theta}_{0} } &~  \left( \tilde{T} - t_{0,2} \right) \log \left(  1 + \frac{ c + t_{0,2} s  }{  \tilde{T} - t_{0,2}  } \right)  \label{eq:Ori_problem_HTT_TS_special8_obj} \\
s.t. &~ 0 \leq t_{0,2} <T, ~\eqref{eq:phase1} 
\end{align}
\end{subequations}
Problem \eqref{eq:Ori_problem_HTT_TS_special8} includes two variables, i.e., $ t_{0,2} $ and $ \bm{\Theta}_{0} $, which can be optimally derived with the closed-form expressions. To solve \eqref{eq:Ori_problem_HTT_TS_special8}, we first fix $ t_{0,2} $ to derive the optimal closed-form solution of $ \bm{\Theta}_{0} $. Similar to \emph{Theorem} \ref{lemma:EQphaseshifts}, the maximization of \eqref{eq:Ori_problem_HTT_TS_special8} with respect to $ \bm{\Theta}_{0} $ is equivalent to maximize the following problem
\begin{align}\label{eq:Optimal_phase1}
\max_{ \bm{\Theta}_{0} }  ~ \sum_{k = 1}^{K} \tilde{c}_{k} \left | \mathbf{g}_{0} \bm{\Theta}_{0} \mathbf{g}_{r,k} + g_{d,k} \right |^{2},~
s.t. ~ \eqref{eq:phase1}.
\end{align}
With some mathematical manipulations in the objective function of \eqref{eq:Optimal_phase1}, we have
\begin{align}
\sum_{k = 1}^{K} \tilde{c}_{k} \left | \mathbf{g}_{0} \bm{\Theta}_{0} \mathbf{g}_{r,k} + g_{d,k} \right |^{2} = \sum_{k = 1}^{K} \tilde{c}_{k}   \left | \bm{\theta}_{0} \mathbf{a}_{k} + g_{d,k} \right |^{2},
\end{align}
where $ \bm{\theta}_{0} = [\exp(j \alpha_{0,1}),..., \exp(j \alpha_{0,N_{R}})] $, and $ \mathbf{a}_{k} = \textrm{diag}(\mathbf{g}_{0})\mathbf{g}_{r,k} $. Thus, \eqref{eq:Optimal_phase1} is equivalent to
\begin{align}\label{eq:Optimal_phase3}
\max_{ \bm{\theta}_{0} } &~ \sum_{k = 1}^{K} \tilde{c}_{k}   \left | \bm{\theta}_{0} \mathbf{a}_{k} + g_{d,k} \right |^{2}  \nonumber\\
s.t. &~ \left | \bm{\theta}_{0} (n) \right | = 1, ~\alpha_{0,n} \in [0,2\pi], ~\forall n \in [1,N_{R}].
\end{align}
To solve \eqref{eq:Optimal_phase3}, we further expand its objective function as 
\begin{align}\label{eq:Obj_subproblem}
&\! \! \!  \sum_{k = 1}^{K} \tilde{c}_{k}  | \bm{\theta}_{0} \mathbf{a}_{k} + g_{d,k} |^{2} 
 = \bm{\theta}_{0} \bm{\Phi}_{1} \bm{\theta}_{0}^{H} + 2 \mathcal{R} \left\{ \bm{\theta}_{0} \bm{\gamma} \right\} + d_{1}.
\end{align}
where 
$
\bm{\Phi}_{1}  = \sum_{k=1}^{K} \tilde{c}_{k} \mathbf{a}_{k} \mathbf{a}_{k}^{H},~ \bm{\gamma} = \sum_{k=1}^{K} \tilde{c}_{k} \textrm{conj} \left( g_{d,k} \right) \mathbf{a}_{k}, \nonumber\\ d_{1}  = \sum_{k=1}^{K} \tilde{c}_{k} g_{d,k} \textrm{conj} \left( g_{d,k} \right). $ 
Substitute \eqref{eq:Obj_subproblem} into \eqref{eq:Optimal_phase3}, we have
\begin{subequations}\label{eq:Subproblem_LC_min}
	\begin{align}
	\min_{\bm{\theta}_{0}} &~  \bm{\theta}_{0} \bm{\Phi} \bm{\theta}_{0}^{H} - 2 \mathcal{R} \left\{ \bm{\theta}_{0}  \bm{\gamma} \right\}  + d   \label{eq:Subproblem_LC_min_obj}\\
	s.t.&~ | \bm{\theta}_{0}(n)| = 1,  ~ \forall n \in [1,N_{R}]. \label{eq:Subproblem_LC_min_constraint}
	\end{align}
\end{subequations}
where $  \bm{\Phi} = -\bm{\Phi}_{1}  $ and $ d = -d_{1} $.
Problem \eqref{eq:Subproblem_LC_min} is still intractable due to its unit modulus equality constraint \eqref{eq:Subproblem_LC_min_constraint}. To solve \eqref{eq:Subproblem_LC_min}, the MM algorithm is adopted to approximate its objective function and feasible set, which iteratively updates the approximated solution to \eqref{eq:Subproblem_LC_min} via an alternating algorithm \cite{Palomar_TSP2017}. 
To perform the MM algorithm, we first investigate the following problem 
\begin{align}\label{eq:MM_problem}
\min_{\mathbf{x}} &~ f_{0}(\mathbf{x}),~
s.t.~ f_{i}(\mathbf{x}) \leq 0, ~ x \in \mathcal{X}, ~i = 1, ..., L,
\end{align} 
where $ f_{i}(x): \mathcal{X} \rightarrow \mathbb{R} $ denotes a continuous function, $ \mathcal{X} $ represents a non-empty closed set.
We approximate both objective function and feasible constraints set of problem \eqref{eq:MM_problem} at each iteration.\footnote{Here we assume that $ f_{i} $ is differential \cite{Palomar_TSP2017}.} Thus, the following convex sub-problem can be solved at the $ m $-th iteration. 
\begin{align}\label{eq:Subproblem_MM}
\min_{\mathbf{x}}&~ g_{0} (\mathbf{x}|\mathbf{x}^{(m)}) ,~
s.t.~ g_{i} (\mathbf{x}|\mathbf{x}^{(m)}) \leq 0, i = 1, ..., L,
\end{align} 
where $ g_{i} (\mathbf{*}|\mathbf{x}^{(m)}) $, $ \forall m = 0, ..., L $ denotes a continuous surrogate function which guarantees the following conditions: 
\begin{align}\label{eq:MM_three_conditions}
g_{i}(\mathbf{x}^{(m)}|\mathbf{x}^{(m)})& = f_{i}(\mathbf{x}^{(m)}), \nonumber\\
g_{i}(\mathbf{x}|\mathbf{x}^{(m)}) & \geq f_{i}(\mathbf{x}), \\
\nabla g_{i} (\mathbf{x}^{(m)}|\mathbf{x}^{(m)}) & = \nabla f_{i} (\mathbf{x}^{(m)}). \nonumber
\end{align}
From \eqref{eq:MM_three_conditions}, it is easily observed that the sequence $ \mathbf{x}^{(m)} $ obtained from \eqref{eq:Subproblem_MM} at each iteration leads to a monogenically decreasing function, i.e., $ f_{0}(\mathbf{x}^{(m)}),~m = 1, 2,...$ which converges to a KKT point \cite{Palomar_TSP2017}. The first and third conditions denote that the surrogate function $ g_{i}(\mathbf{x}^{(m)}|\mathbf{x}^{(m)}) $ and its first-order approximation are the same as the the original function $ f_{i}(\mathbf{x}^{(m)}) $ and its first-order approximation at $ \mathbf{x}^{(m)} $. The second condition implies that the surrogate function $ g_{i}(\mathbf{x}|\mathbf{x}^{(m)}) $ is constructed based on the upper bound of the original function $ f_{i}(\mathbf{x}^{(m)}) $. To perform the MM algorithm, it is imperative to achieve the surrogate function $ g_{i}(\mathbf{x}|\mathbf{x}^{(m)}) $, guaranteeing these conditions \eqref{eq:MM_three_conditions} and better tractability than $ f_{i}(\mathbf{x}^{(m)}) $ \cite{Palomar_TSP2017,CPAN_MCELL_2019}. 

The following \emph{theorem} is required to derive the optimal closed-form solution of $ \bm{\theta}_{0} $ by the MM algorithm.
\begin{theorem}\label{theorem:Optimal_phase_WET}
	The optimal solution to problem \eqref{eq:Subproblem_LC_min} is derived in terms of the following closed-form expression via the MM algorithm.
\begin{align}\label{eq:Closed_form_theta0}
\bm{\theta}_{0}^{*} = \left[\!\!\begin{array}{ccc}
\exp \left(j \arg [\tilde{\bm{\gamma}}(1)] \right) \!\!& ,..., & \!\!\exp \left(j \arg [\tilde{\bm{\gamma}}(N_{R})] \right)
\end{array}\!\!\right],
\end{align}
where  $ \tilde{\bm{\gamma}} =   \left( \lambda_{ \max } ( \bm{\Phi} ) \mathbf{I}_{N_{R} \times N_{R}} - \bm{\Phi} \right) \tilde{\bm{\theta}}_{0}^{H} + \bm{\gamma}  $; $ \lambda_{\max} ( \bm{\Phi} ) $ is the maximum eigenvalue of $ \bm{\Phi} $; $ \tilde{\bm{\theta}}_{0} $ denotes the approximated solution to $ \bm{\theta}_{0} $ which is achieved in the previous iteration of the alternating algorithm. 
\end{theorem}
\begin{IEEEproof}
	See Appendix \ref{appendix:Optimal_phase_WET}.
	\end{IEEEproof}
\begin{remark}
	In \eqref{eq:Closed_form_theta0}, we obtain the optimal phase shift of the WET phase, which can be independently applied to the IRS to maximize the energy signal strength at sensor nodes. 
	Although problem \eqref{eq:Optimal_phase3} can be easily relaxed as an SDP, \eqref{eq:Closed_form_theta0} provides an optimal solution which is more efficient for implementation and significantly reduces the computational complexity introduced by the SDP, especially for the larger reflecting elements $ N_{R} $.
\end{remark}
Since the optimal solution of $ \bm{\theta}_{0} $ can be iteratively updated via the approximation of the MM algorithm, its convergence property is characterized as follows. The MM algorithm shows that the objective value $ f(\bm{\theta}_{0}) $ is monotonically non-increasing at each iteration, i.e., 
	\begin{align}\label{eq:MM_objective}
	\!\!\!\!\!\!	f(\bm{\theta}_{0}^{(m+1)}) \! \leq \! g(\bm{\theta}_{0}^{(m+1)}|\bm{\theta}_{0}^{(m)}) \! \leq \! g(\bm{\theta}_{0}^{(m)}|\bm{\theta}_{0}^{(m)}) \!= \! 	f(\bm{\theta}_{0}^{m}),\!\!\!\!
	\end{align}
	where $ \bm{\theta}_{0}^{(m)} $ denotes the point generated by the MM algorithm at the $ m $-th iteration. From \eqref{eq:MM_objective}, the first inequality and the third equality follow the first two relations of \eqref{eq:MM_three_conditions}, respectively. Also the second inequality holds via solving \begin{align}
	\!\!\!\!\! \bm{\theta}_{0}^{*} \!=\! \arg \min_{ \bm{\theta}_{0}(n)  }~ g(\bm{\theta}_{0}|\bm{\tilde{\theta}}_{0} ), ~
	s.t.~ | \bm{\theta}_{0}(n) | \!=\! 1,~\forall n \in [1,N_{R}].
	\end{align} 
	The monotonicity of the objective function in the above problem confirms that the MM algorithm converges to a stationary point in practice due to the unit modulus equality constraint \cite{Palomar_TSP_2016,Palomar_TSP2017}.

By exploiting \emph{Theorem} \ref{theorem:Optimal_phase_WET}, the optimal phase shift matrix of the WET, i.e., $ \bm{\Theta}_{0}^{*} $, can be achieved from $ \bm{\theta}_{0}^{*} $.
For a given $ \bm{\Theta}_{0}^{*} $, problem \eqref{eq:Ori_problem_HTT_TS_special8} is a single-variable optimization problem which can easily be proved to be convex with respect to $ t_{0,2} $. Thus, it can be solved via one-dimensional line search, i.e., golden search method. Here, we derive the optimal second sub-slot of the WET phase in terms of closed-form expression via the following \emph{theorem}:
\begin{theorem}\label{theorem:Optimal_t02}
	The optimal second sub-slot of the WET phase, i.e., $ t_{0,2}^{*} $ is derived as
\begin{align}\label{eq:Optimal_second_subslot}
t_{0,2}^{*} = \frac{  \left\{  \exp \left[ \mathcal{W} \left( \frac{ s - 1 }{ \exp(1) }  \right) + 1 \right] - 1  \right\} \tilde{T}  - c   }{ s + \exp \left[ \mathcal{W} \left( \frac{ s - 1 }{ \exp(1) }  \right) + 1 \right] - 1  }.
\end{align}
\end{theorem}
\begin{IEEEproof}
	See Appendix \ref{appendix:Optimal_t02}.
	\end{IEEEproof}
The proposed low complexity algorithm to solve problem \eqref{eq:Ori_problem_HTT_TS} is summarized in \emph{Algorithm} \ref{algorithm:LC_algorithm}. Now, we characterize the computational complexity of \emph{Algorithm} \ref{algorithm:LC_algorithm}, which depends upon the proposed MM algorithm. Thus, the total computational complexity of the MM algorithm is given as $ \mathcal{O} \left( N_{R}^{3} + I_{\max} N_{R}^{2} \right) $, where $ I_{\max} $ is the number of iterations the MM algorithm to achieve convergence. 
\vspace{0.5em}
\hrule
\vspace{0.5em}
\begin{algorithm}\label{algorithm:LC_algorithm}
	Proposed algorithm for the special case ($ P_{c,k} = 0 $) to solve problem \eqref{eq:Ori_problem_HTT_TS}.
	\vspace{0.5em}
	\hrule
	\vspace{0.5em}
	\begin{enumerate}
		\item \textbf{Input}: $ P_{0} $, $ \eta $, and $ c_{k},~\forall k \in [1,K] $.
		\item \textbf{Obtain} the optimal phase shifts of the WIT phase $ \bm{\Theta}_{k},~\forall k \in [1,K] $ according to $ \bm{\theta}_{k}^{*} $ in \eqref{eq:Optimal_phaseWIT}.
		\item \textbf{Obtain} the first sub-slot of the WET phase $ t_{0,1}^{*} $ via \eqref{eq:Optimal_t01} in \emph{Lemma} \ref{lemma:Handle1}.
		\item \textbf{Obtain} the optimal phase shift of the WET phase $ \bm{\Theta}_{0}^{*} $, which is from $ \bm{\theta}_{0}^{*} $ in \eqref{eq:Closed_form_theta0} via the MM algorithm.
		\item \textbf{Obtain} the second sub-slot of the WET phase $ t_{0,2}^{*} $ via \eqref{eq:Optimal_second_subslot} in \emph{Theorem} \ref{theorem:Optimal_t02}.
		\item \textbf{Substitute} $ \bm{\Theta}^{*} $ and $ t_{0,i}^{*}, ~\forall i \in [1,2] $ into \eqref{eq:tauk1} to obtain the optimal time allocation of the WIT phase $ \tau_{k}^{*} $, $ k \in [1,K] $.
	   \item \textbf{Substitute} $ \bm{\Theta}^{*} $, $ t_{0,i}^{*} $, and $ \tau_{k} $ into \eqref{eq:Optimal_Pk0} in \emph{Lemma} \ref{lemma:Handle1} to obtain the optimal transmit power of each IoT sensor $ P_{k},~\forall k \in [1,K] $.
		\item \textbf{Output}: $ t_{0,i}^{*},~\forall i \in [1,2] $, $ \tau_{k}^{*},~\forall k \in [0,K] $, $ \bm{\Theta}_{k}^{*},~\forall k \in [0,K] $, and $ P_{k}^{*},~\forall k \in [1,K] $.
	\end{enumerate}
\vspace{0.5em}
	\hrule
\end{algorithm}
\vspace{0.5em}

\subsubsection{General Case ($ P_{c,k} \neq 0 $)}\label{section:TS_general}
In this subsection, we consider the general case of $ P_{c,k} \neq 0 $, which is more practical and complex than the special case studied in Section \ref{section:Sepecial_TS}. Here, we write problem \eqref{eq:Ori_problem_HTT_TS} as 
\begin{subequations}\label{eq:Ori_problem_HTT_TS_general0}
	\begin{align}
	\max_{\bm{\Omega}_{\textrm{IRS-HTT-TS}}} &~  \sum_{k=1}^{K} \tau_{k} \log \left( 1 + \frac{ P_{k} c_{k} }{ \sigma^{2} }  \right), \label{eq:Ori_problem_HTT_TS_general0_obj} \\ 
	s.t. &~ \eqref{eq:HTT_TS_sensorpower},~\eqref{eq:HTT_TS_IRSpower},~\eqref{eq:HTT_TS_time},~\eqref{eq:HTT_TS_timeallocation0},~
	\eqref{eq:phase1},~\eqref{eq:HTT_TS_variables1},~\eqref{eq:HTT_TS_variables2}, 
	\end{align}
\end{subequations}

Due to the non-convexity of problem \eqref{eq:Ori_problem_HTT_TS_general0}, we introduce the SDP relaxation to solve it. To proceed, we define $ \tilde{P}_{k} = \tau_{k} P_{k} $, $ \mathbf{W}_{0} = t_{0,2} \mathbf{V}_{0} $, $ \mathbf{V}_{0} = \tilde{\bm{\theta}}_{0}^{H} \tilde{\bm{\theta}}_{0} $, $ \tilde{\bm{\theta}}_{0} =  \left[\!\!\begin{array}{cc}
\bm{\theta}_{0} \!&\! 1
\end{array}\!\!\right] $, $ \mathbf{\tilde{a}}_{k} = \left[\!\! \begin{array}{cc}
\mathbf{a}_{k} \! \\  \! g_{d,k}
\end{array} \!\!\right] $, $ \mathbf{A}_{k} = \mathbf{\tilde{a}}_{k} \mathbf{\tilde{a}}_{k}^{H} $, and $ \mathbf{a}_{k} = \textrm{diag}(\mathbf{g}_{0}) \mathbf{g}_{r,k}  $ to handle the objective function \eqref{eq:Ori_problem_HTT_TS_general0_obj}  and the constraint \eqref{eq:HTT_TS_sensorpower}, respectively. In addition, the optimal first sub-slot of the WET phase $ t_{0,1}^{*} $ can be derived as \eqref{eq:Optimal_t01} in \emph{Lemma} \ref{lemma:Handle1}. Thus, problem \eqref{eq:Ori_problem_HTT_TS_general0} is relaxed as 
\begin{subequations}\label{eq:TS_SDP}
	\begin{align}
	&~	\max_{\bm{\Omega}_{\textrm{IRS-HTT-TS}}}   \sum_{k=1}^{K} \tau_{k} \log \left( 1 + \frac{ \tilde{P}_{k} c_{k} }{ \tau_{k} \sigma^{2} }  \right),  \\ 
	&	s.t. ~ \tilde{P}_{k} \!+\!  \tau_{k}  P_{c,k}  \! \leq \! 
		 t_{0,1} \eta P_{0} \left | g_{d,k} \right |^{2} \!+ \! \eta P_{0} \textrm{Tr}\left( \mathbf{A}_{k} \mathbf{W}_{0} \right), \\
		&~~~~~ N_{R} P_{c,\textrm{IRS}} \left( t_{0,2} + \sum_{k = 1}^{K} \tau_{k} \right) \leq 
		t_{0,1}^{*} \eta P_{0} \left \| \mathbf{g}_{0} \right \|^{2}, \\
		&~~~~~ t_{0,1}^{*} + t_{0,2} + \sum_{k=1}^{K} \tau_{k} \leq T, \\
		&~~~~~ \mathbf{W}_{0}(n,n) = t_{0,2}, ~\forall n \in [1,N_{R}+1], \\
		&~~~~~ \mathbf{W}_{0} \succeq \mathbf{0}, \\
		&~~~~~ \textrm{rank}(\mathbf{W}_{0}) = 1, \label{eq:Nonconvex_rankone}\\
		&~~~~~ t_{0,2} \geq 0,~\tau_{k} \geq 0, ~\forall k \in [0,K],\nonumber\\ &~~~~~~~~\tilde{P}_{k} \geq 0, ~\forall k \in [1,K], \\
		&~~~~~ \bm{\Omega}_{\textrm{IRS-HTT-TS}} = \left[ \mathbf{W}_{0}, t_{0,2}, \left\{ \tau_{k} \right\}_{1}^{K}, \left\{ \tilde{P}_{k} \right\}_{1}^{K}   \right]. 
	\end{align}
\end{subequations}
By relaxing the non-convex rank-one constraint \eqref{eq:Nonconvex_rankone}, it can be easily verified that \eqref{eq:TS_SDP} is a convex optimization problem, which can be solved by the interior-point methods \cite{boyd_B04}. The optimal solution of problem \eqref{eq:TS_SDP} is represented by $ \left( \mathbf{W}_{0}^{*}, t_{0,2}^{*}, \{ \tau_{k}^{*} \}_{1}^{K}, \{ \tilde{P}_{k}^{*} \}_{1}^{K} \right) $, and we further obtain the optimal solution $ \mathbf{V}_{0}^{*} = \frac{\mathbf{W}_{0}^{*}}{\tau_{0}^{*}} $ and $ P_{k} = \frac{\tilde{P}_{k}}{ \tau_{k}^{*} } $. Note that the relaxed problem \eqref{eq:TS_SDP} may incur a higher rank solution, i.e., $ \textrm{rank}(\mathbf{V}_{0}^{*}) > 1 $. To circumvent this issue, we apply eigenvalue decomposition on $ \mathbf{V}_{0}^{*}  $, yielding, $ \mathbf{V}_{0}^{*} = \bm{\Upsilon} \bm{\Lambda} \bm{\Upsilon}^{H} $, where $ \bm{\Upsilon} \in \mathbb{C}^{(N_{R}+1) \times (N_{R}+1)} $ is a
unitary matrix, and $ \bm{\Lambda} \in \mathbb{C}^{(N_{R}+1) \times (N_{R}+1)} $ is a diagonal matrix with eigenvalues arranged in decreasing order. To proceed, we construct a suboptimal solution as $ \tilde{\bm{\theta}}_{0}^{H} = \bm{\Upsilon} \bm{\Lambda}^{\frac{1}{2}} \bm{\kappa} $, where  $ \bm{\kappa} \in \mathbb{C}^{(N_{R}+1) \times 1} $  
is randomly generated to follow complex
circularly symmetric uncorrelated Gaussian distribution
with zero-mean and  covariance matrix $ \mathbf{I}_{(N_{R}+1)\times (N_{R}+1)} $. Thus, we obtain the optimal phase shift vector  $ \bm{\theta}_{0}^{*}(n) = \exp \left( j \arg\left( \tilde{\bm{\theta}}_{0}(n) \right) \right) $ for $ n \in [1,N_{R}] $ and the optimal phase shift matrix $ \bm{\Theta}_{0}^{*} $ can be achieved from $ \bm{\theta}_{0}^{*} $ \cite{QWu_TWC_2019}. 
Since \eqref{eq:Ori_problem_HTT_TS_general0} can be relaxed to a convex problem that satisfies Slater's condition, the strong duality holds. To gain more insights, we characterize the optimal closed-form solution of the time allocation, i.e., $ t_{0,i},~\forall i \in [1,2] $ and $ \tau_{k}, ~\forall k \in [1,K] $ for given phase shifts $ \bm{\Theta}_{k}^{*},~\forall k \in [0,K] $. 
Note that the optimal first sub-slot of the WET phase $ t_{0,1}^{*} $ has been derived as \eqref{eq:Optimal_t01} in \emph{Lemma} \ref{lemma:Handle1}.
To proceed, we modify \eqref{eq:Ori_problem_HTT_TS_general0} to an equivalent form of
\begin{subequations}\label{eq:HTT_TS_general_time}
\begin{align}
\max_{ t_{0,2}, \tau_{k} } &~  \sum_{k=1}^{K} \tau_{k} \log \left( 1 + \frac{ a_{k}+ t_{0,2} b_{k}     }{ \tau_{k} }  - d_{k}  \right),  \\ 
s.t.  
&~ t_{0,1}^{*} + t_{0,2} + \sum_{k=1}^{K} \tau_{k} \leq T, \label{eq:Timeconstraint} \\
&~ t_{0,2} \geq 0, ~\tau_{k} \geq 0,~\forall k \in [1,K].
\end{align}
\end{subequations}
where $ a_{k} = \bar{c}_{k} \tilde{c}_{k} $, $ b_{k} = s_{k}^{*} \tilde{c}_{k} $,  $ d_{k} = P_{c,k} \hat{c}_{k} $, $ \hat{c}_{k} = \frac{c_{k}}{\sigma^{2}} $, and $ s_{k}^{*} = \left | \mathbf{g}_{0} \bm{\Theta}_{0}^{*} \mathbf{g}_{r,k} \!+\! g_{d,k} \right |^{2} $. 
To derive the optimal solution of problem \eqref{eq:HTT_TS_general_time}, the following \emph{theorem} is in order.
\begin{theorem}\label{theorem:Optimal_generalcase1}
	The optimal closed-form time slots of problem \eqref{eq:HTT_TS_general_time}, e.g., $ \tilde{t}_{0,2}^{*} $ and $ \tilde{\tau}_{k},~\forall k \in [1,K] $, are derived, respectively, as 
		\begin{align}
		\tilde{t}_{0,2}^{*}  = \frac{T - t_{0,1}^{*} - \sum_{k=1}^{K} \frac{ a_{k} }{ x_{k}^{*} } }{ 1 + \sum_{k=1}^{K} \frac{ b_{k} }{ x_{k}^{*} } }, ~
    	\tilde{\tau}_{k}^{*}  = \frac{ a_{k} + t_{0,2}^{*} b_{k} }{ x_{k}^{*} }. \label{eq:tauk11}
		\end{align}
\end{theorem}
\begin{IEEEproof}
	See Appendix \ref{appendix:Optimal_generalcase1}.
	\end{IEEEproof}
\emph{Theorem} \ref{theorem:Optimal_generalcase1} unveils an insight that when the phase shifts of the WET and WIT stages are given, we can calculate the optimal time slots $ t_{0,1}^{*} $, $ \tilde{t}_{0,2}^{*} $, and $ \tilde{\tau}_{k}^{*} $ to obtain the maximum achievable sum throughput. 
\subsection{Discrete Phase Shift Optimization}
In Section \ref{section:Phase_WIT} and Section \ref{section:Optimal_IRS_HTT_TS}, we obtained the optimal continuous phase shifts for the WIT and WET phases. In this subsection, we obtain the quantized phase shifts according to constraint \eqref{eq:Dis_phaseshift_constraint}. We denote the optimal continuous phase shifts as $ \bm{\tilde{\theta}}^{*} = \{\bm{\tilde{\theta}}_{0}^{*}, ..., \bm{\tilde{\theta}}_{K}^{*} \} $, where $ \bm{\tilde{\theta}}_{k}^{*} = [\tilde{\theta}_{k,1}^{*}, ..., \tilde{\theta}_{k,N_{R}}^{*}],~\forall k \in [0,K] $ is derived in Section \ref{section:Phase_WIT} and Section \ref{section:Optimal_IRS_HTT_TS}, and substituted into \eqref{eq:Dis_phaseshift_constraint}. Thus, the optimal discrete phase shift of the $ n $-element at the $ k $-time slot, denoted by $ \bar{\theta}_{k,n}^{*} $ is derived as 
\begin{align}\label{eq:Optimal_dis}
\bar{\theta}_{k,n}^{*} = \exp( j \alpha_{l^{*}} ),
\end{align}
where $ l^{*} \in [1,L] $ denotes the optimal index of discrete phase shift set $ \mathcal{S}_{\bm{\Theta}} $, which can be determined by solving the following problem
\begin{align}\label{eq:Optimal_index}
l^{*} = \arg \min_{ l \in [1, L]} \left | \tilde{\theta}_{k,n}^{*} -  \exp( j \alpha_{l} ) \right|.
\end{align}
Note that the discrete optimal phase shifts of the IRS can be obtained by first solving \eqref{eq:Ori_problem_HTT_TS} with the continuous phase shifts instead of \eqref{eq:Ori_problem_HTT_TS0}, and then quantizing the continuous phase shifts to their nearest points from $ \mathcal{S}_{\bm{\Theta}} $ via \eqref{eq:Optimal_dis}.
\section{Numerical Results}\label{section:Numerical}
\begin{figure}[!htbp]
	\centering
	\includegraphics[scale = 0.4]{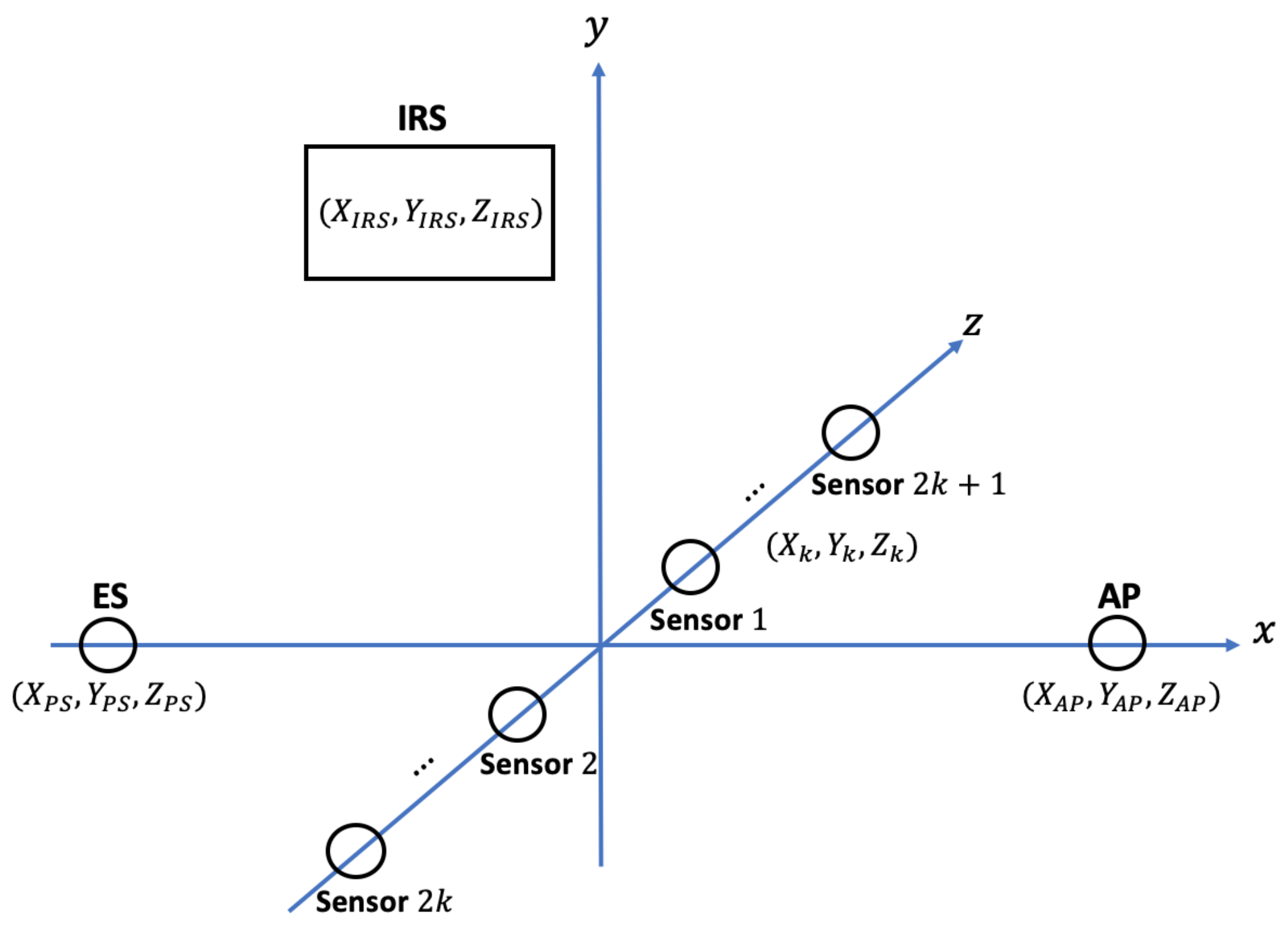}
	\caption{System deployment.}
	\label{fig:System_deployment}
\end{figure}
This section provides numerical evaluations to validate the theoretical derivations of the proposed IRS-HTT-TS policy in Section \ref{section:IRSHTTTSPS}. In our simulations, we use a three-dimensional (3D) coordinate in Fig. \ref{fig:System_deployment} showing the possible deployment of the IRS assisted WPSN. Specifically, we assume that the coordinates of the ES, the AP, and the IRS are located at $ (-10,0,0) $, $ (10,0,0) $ and $ (-2,6,0) $, respectively, while the sensor nodes are assigned at $ \left( 0,0, \frac{l \times d_{I} }{2} \right) $ if $ l = 1,...,(2k+1) $, and $ \left(0,0,-\frac{(l -1)\times d_{I} }{2} \right) $, if $ l = 2,...,2k $, where $ d_{I} $ is the separation distance between two neighbouring sensors. The channel coefficient is composed of distance-dependent path loss model and small-scale fading. The path loss model is set to $ P_{L} = A d^{ -\varepsilon } $, where $ A = -20~\textrm{dB} $, $ \varepsilon $ is the path loss exponent, and $ d $ represents the distance between any two devices, i.e., the ES and the IRS, the ES and the $ k $-th sensor node, the IRS and the $ k $-th sensor node, the IRS and the AP, as well as the $ k $-th sensor node and the AP.
The channel coefficients between the ES and the IRS, the IRS and the $ k $-th sensor node, the $ k $-th sensor node and the IRS, as well as the IRS and the AP are modelled as $ \mathbf{g}_{0} = \sqrt{ \frac{K_{1}}{K_{1}+1} } \mathbf{g}_{0}^{\textrm{LOS}} + \sqrt{\frac{1}{K_{1}+1}} \mathbf{g}_{0}^{\textrm{NLOS}} $, $ \mathbf{g}_{r,k} = \sqrt{ \frac{K_{1}}{K_{1}+1} }  \mathbf{g}_{r,k}^{\textrm{LOS}} +  \sqrt{ \frac{ 1 }{ K_{1}+1} } \mathbf{g}_{r,k}^{\textrm{NLOS}} $,
$ \mathbf{h}_{k} = \mathbf{g}_{r,k}^{T} $, 
and $ \mathbf{h}_{r} = \sqrt{ \frac{K_{1}}{K_{1}+1} } \mathbf{h}_{r}^{\textrm{LOS}} + \sqrt{ \frac{ 1 }{ K_{1}+1} } \mathbf{h}_{r}^{\textrm{NLOS}} $.
Specifically, $ \mathbf{g}_{0}^{\textrm{LOS}} $, $ \mathbf{g}_{r,k}^{\textrm{LOS}} $, and $ \mathbf{h}_{r}^{\textrm{LOS}} $
denote the line-of-sight (LOS) deterministic components of the corresponding channel coefficients. Also, we denote one of the array responses as $ \mathbf{g}_{0}^{\textrm{LOS}} = \left[ 1, \exp \left( - j \frac{2 \pi d }{ \lambda } \phi \right), ..., \exp \left( - j \frac{2 \pi (N_{R} - 1) d }{ \lambda } \phi  \right) \right] $ and others are similarly generated, where $ \lambda $ denotes the carrier wavelength, $ \phi = \cos^{-1} \left( \frac{X_{PS} - X_{IRS}}{d_{PS2IRS}} \right) $ is the angle of arrival(AoA)/angle of departure(AoD); $ \mathbf{g}_{0}^{\textrm{NLOS}} $, $ \mathbf{g}_{r,k}^{\textrm{NLOS}} $, and $ \mathbf{h}_{r}^{\textrm{NLOS}} $ are the non-line-of-sight  (NLOS) components of the corresponding channel coefficients which follow the Rayleigh distribution; $ K_{1} $ is the Rician factor which is set to $ 6~\textrm{dB} $ without loss of generality. The remaining small-scale channel coefficients are generated as circularly symmetric Gaussian random variables with zero mean and unit variance. Other configurations of the simulations are summarized as: the total transmission time period $ T = 1~\textrm{s} $, number of the reflecting elements at the IRS $ N_{R} = 30 $, number of sensor nodes $ K = 6 $, transmit power at the ES $ P_{0} = 30~\textrm{dBm} $, noise power at the AP $ \sigma^{2} = -100~\textrm{dBm} $, energy conversion efficiency $ \eta = 0.8 $, the circuit power of the $ k $-th sensor node as well as the IRS $ P_{c,k}  = P_{c,\textrm{IRS}} = P_{c} = 0.01 ~\textrm{mW} $. The selected configurations and parameters, especially those related to the power and energy, are widely considered in the literature, e.g., \cite{QQWu_TVT_2018}, and the low-power circuit has been justified and supported by the practical investigations, e.g., \cite{7932881}. The number of bits used to indicate the number of of phase resolution $ B = 1~\textrm{or}~2 $, respectively, unless specified. 
We evaluate the performance gains of the proposed low complexity scheme in comparison to the following benchmark schemes:
\begin{enumerate}
	\item \emph{SDP relaxation}: In the special case, problem \eqref{eq:Ori_problem_HTT_TS2} is relaxed as a convex optimization problem, and solved directly by interior-point methods \cite{boyd_B04}. 
	\item \emph{Random phase shift}: The phase shifts are randomly generated, whereas the transmission time slots are optimally designed in Section \ref{section:Optimal_IRS_HTT_TS}. 
	\item \emph{Without IRS}: This system model degrades into the conventional WPSN \cite{Rui_Zhang_WPCN_TWC_2014}, where the transmission time allocations are optimally designed.
	\item \emph{Upper bound}: In this scheme, the IRS does not consider the energy collection to support its circuit operation, i.e., $ P_{c,\textrm{IRS}} = 0 $, which leads to the first sub-slots of the WET phase $ t_{0,1} = 0 $. 
\end{enumerate}
To distinguish different schemes, ``SDP'' and ``SDP: discrete phase shift ($ B = 1~ \textrm{or}~2 $)'' denote the SDP relaxation with continuous and discrete phase shifts, respectively; ``LC'' and ``LC: discrete phase shift ($ B = 1~ \textrm{or}~2 $)'' denote the proposed low complexity scheme with continuous and discrete phase shifts, respectively.

\begin{figure}[!htbp]
	\centering
	\includegraphics[scale = 0.5]{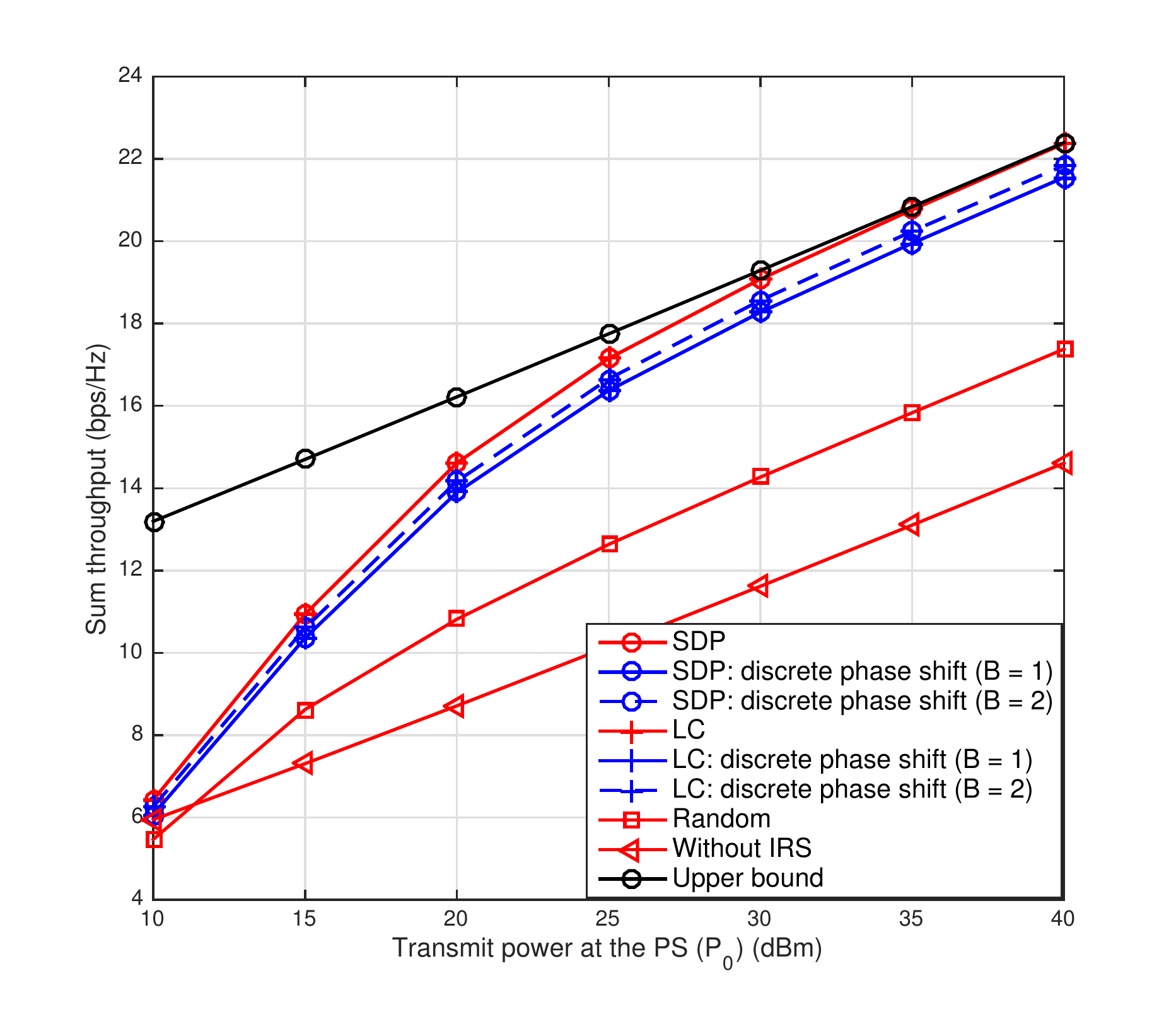}
	\caption{Sum throughput versus transmit power at the ES $ P_{0} $ for special case.}
	\label{fig:Rate_vs_P0_special1}
\end{figure}
\begin{figure}[!htbp]
	\centering
	\includegraphics[scale = 0.5]{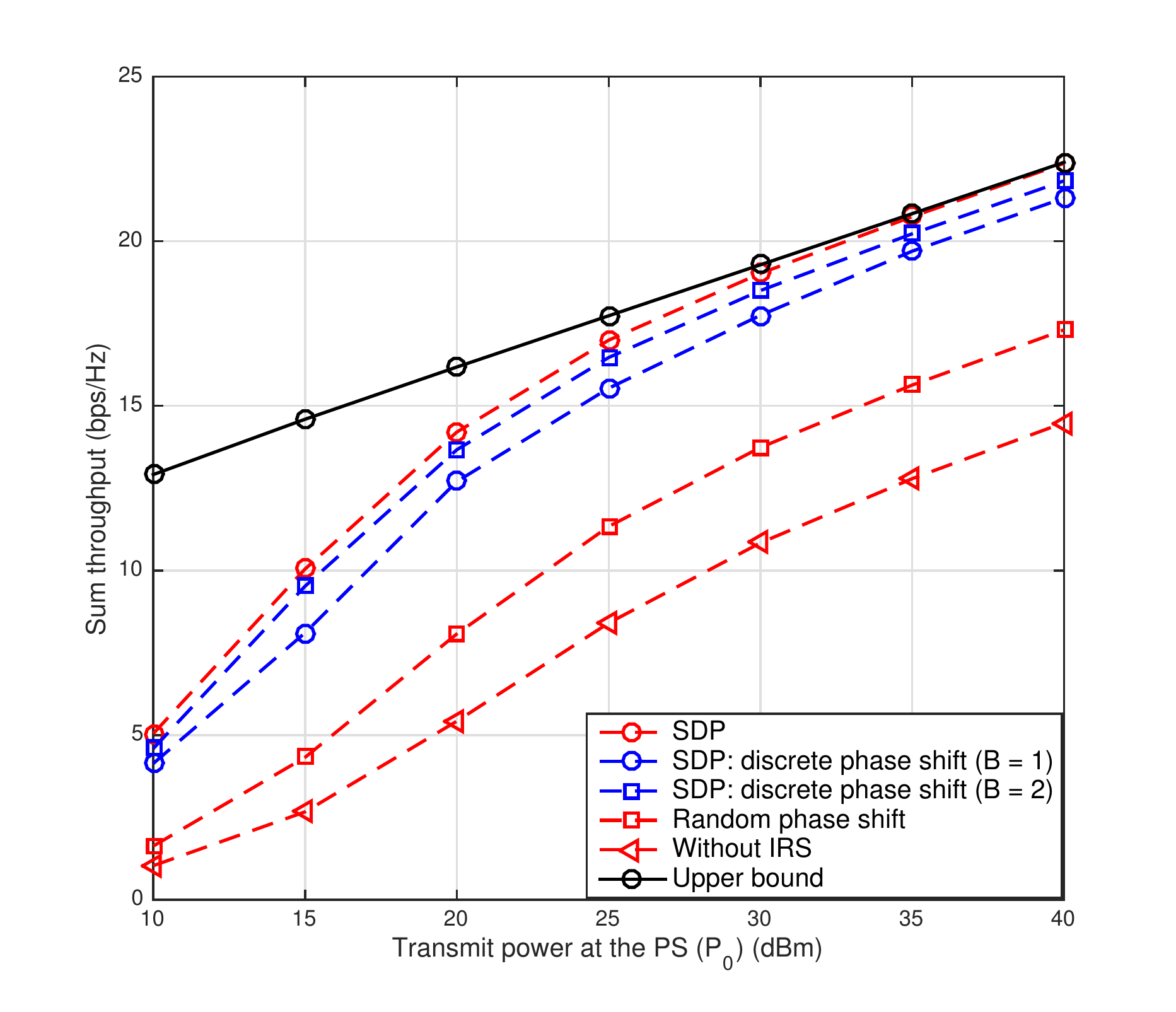}
	\caption{Sum throughput versus transmit power at the ES $ P_{0} $ for general case.}
	\label{fig:Rate_vs_P0_general1}
\end{figure}
First, we evaluate the sum throughput versus the transmit power at the ES $ P_{0} $ for the special and general cases in Fig. \ref{fig:Rate_vs_P0_special1} and Fig. \ref{fig:Rate_vs_P0_general1}. From these results, it is seen that the sum throughput has an increasing trend with $ P_{0} $. 
Also, the scheme with continuous phase shifts slightly outperforms its counterpart with quantized phase shifts $ B = 1~\textrm{or}~2 $. This is due to the fact that the quantized phase shifts leads to the imperfect alignment between the direct and reflected links for the WET and WIT phases and results in a performance loss.
In addition, the quantized phase shift case with $ B = 2 $ has a slightly higher sum throughput than that with $B = 1 $. This is due to the fact that a larger $ B $ brings higher granularity of phase shifts to be selected for energy/information reflection. Moreover, the proposed scheme has an obvious performance gain compared with the schemes with random phase shifts and without IRS, which demonstrates the beneficial role of IRS to enhance the WET and WIT, especially in the large transmit power region. 
From Fig. \ref{fig:Rate_vs_P0_special1}, one can also observe that the proposed low complexity scheme achieves a very close performance with the SDP relaxed counterpart, which verifies the effectiveness of our proposed low complexity scheme derived in Section \ref{section:Optimal_IRS_HTT_TS}. 

\begin{figure}[!htbp]
	\centering
	\includegraphics[scale = 0.5]{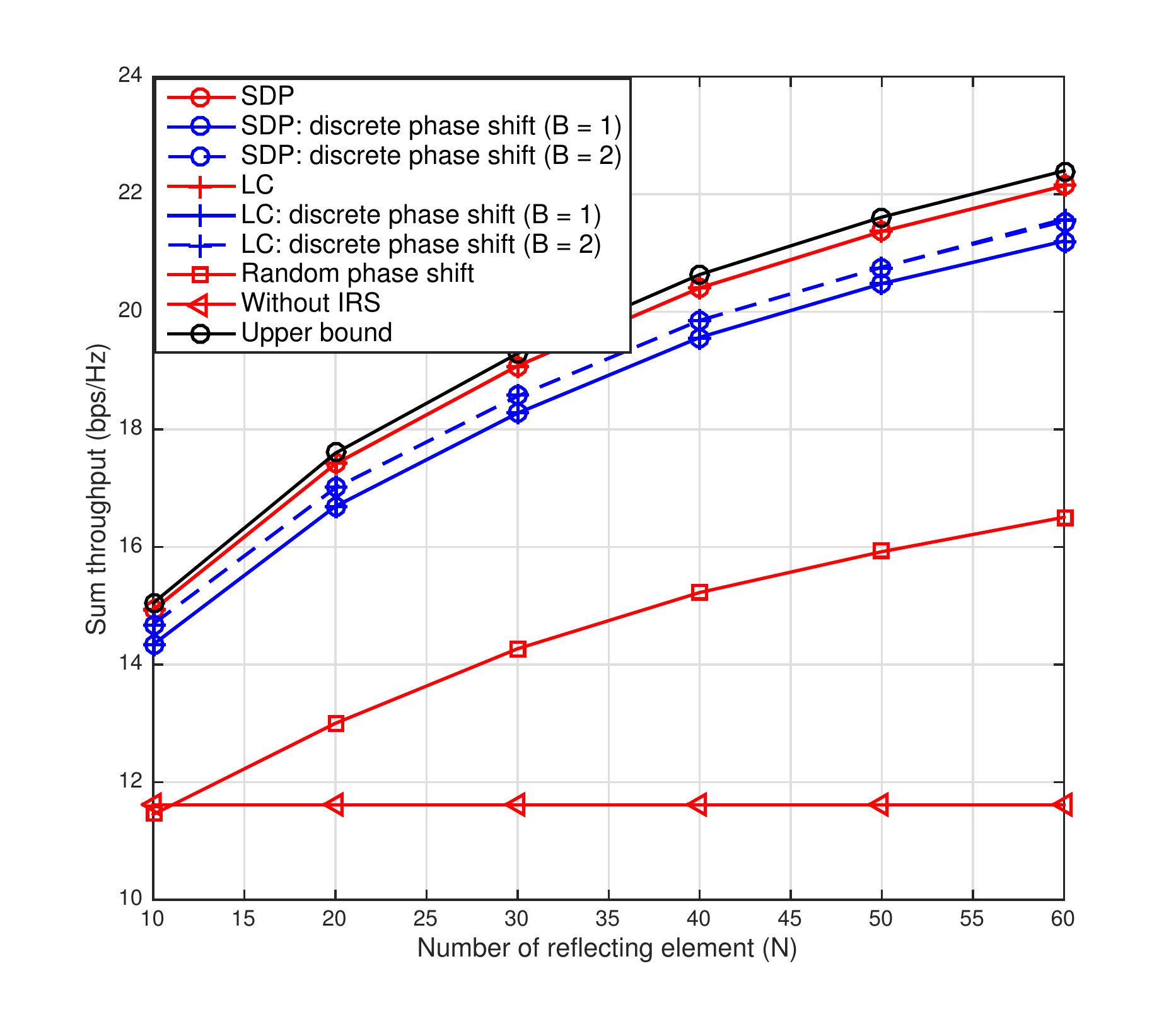}
	\caption{Sum throughput versus number of reflecting elements $ N_{R} $ for special case.}
	\label{fig:Rate_vs_N_R_special}
\end{figure}
\begin{figure}[!htbp]
	\centering
	\includegraphics[scale = 0.5]{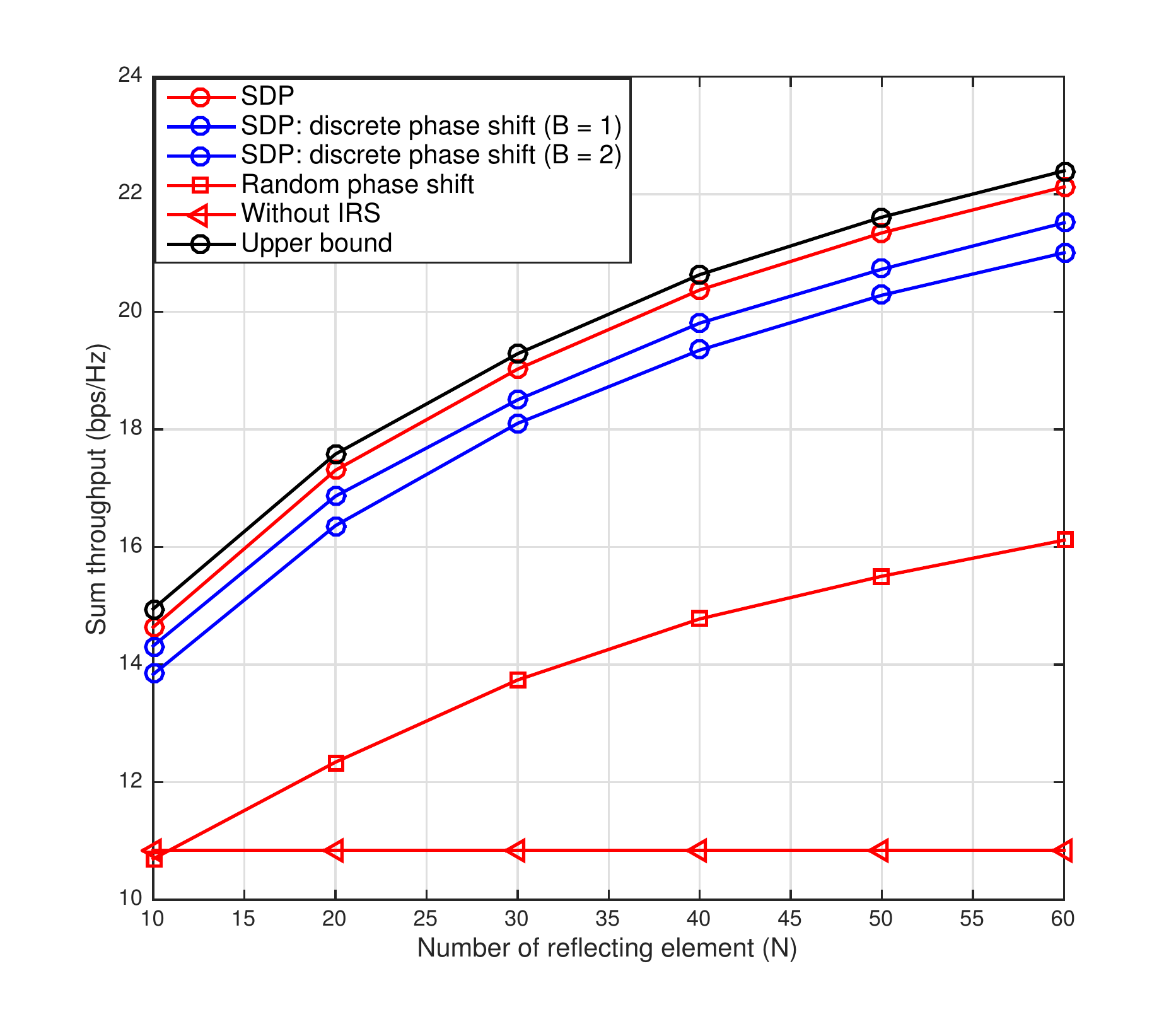}
	\caption{Sum throughput versus number of reflecting elements $ N_{R} $ for general case.}
	\label{fig:Rate_vs_N_R_general}
\end{figure}
Fig. \ref{fig:Rate_vs_N_R_special} and Fig. \ref{fig:Rate_vs_N_R_general} evaluate the impact of the number of reflecting elements $ N $ on the sum throughput. From these results, we see that the proposed scheme can effectively enhance the sum throughput with more reflecting elements in comparison to the schemes with random phase shift and without IRS. Also, the scheme with quantized phase shifts $ B = 1~\textrm{or}~2 $ has a slight performance loss than that with continuous phase shifts. This is because the quantized phase shifts incur imperfect alignment between the direct and reflected links for the WET and WIT phases, resulting in degraded throughput performance. To validate the proposed low complexity scheme, we compare it with the SDP relaxed scheme. By comparison, the two schemes have a comparable performance, which confirms the effectiveness of our proposed low complexity scheme derived in Section \ref{section:Optimal_IRS_HTT_TS}. 

\begin{figure}[!htbp]
	\centering
	\includegraphics[scale = 0.5]{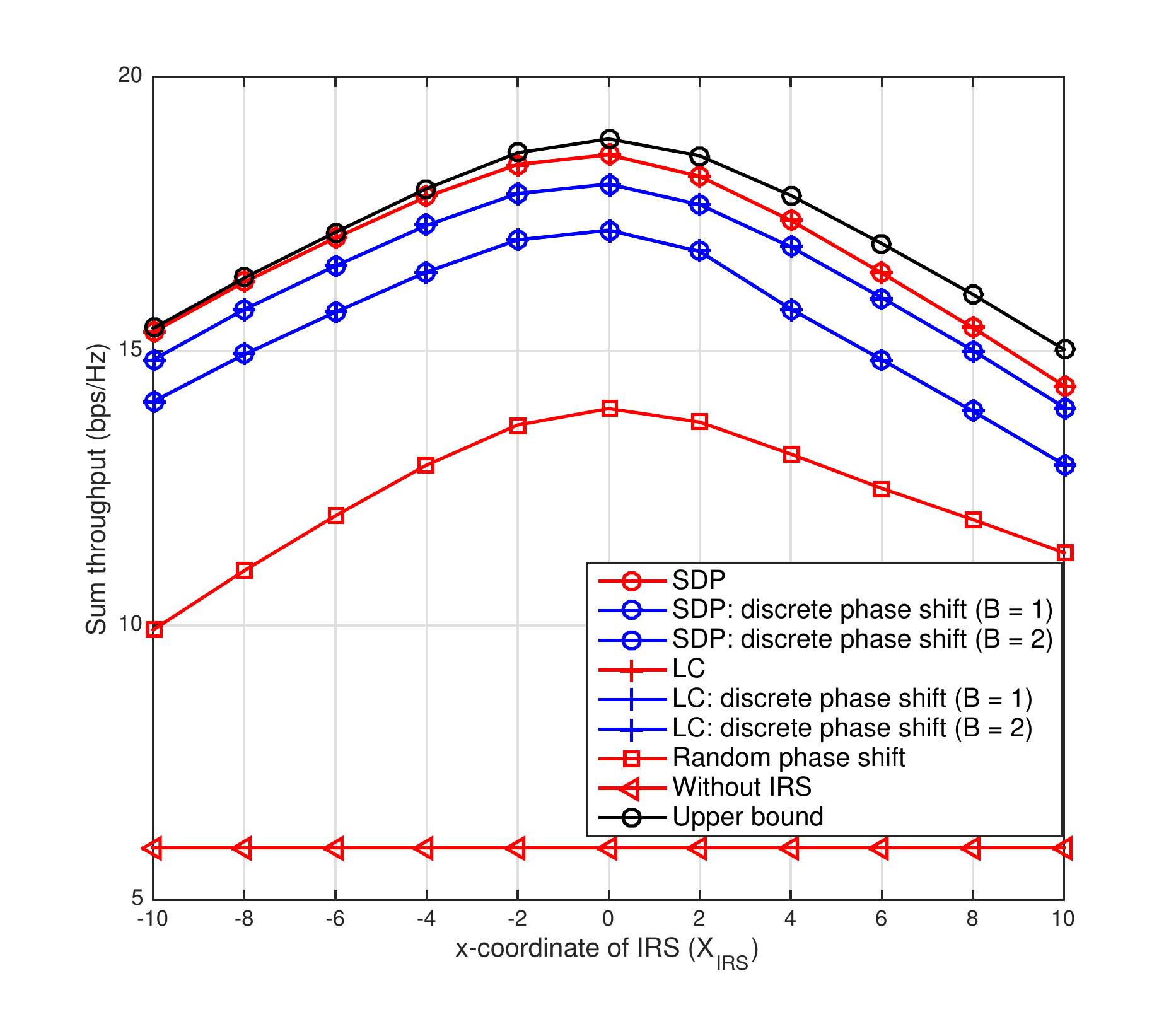}
	\caption{Sum throughput versus x-coordinate of IRS $ X_{\textrm{IRS}} $ for special case.}
	\label{fig:Rate_vs_X_IRS_special}
\end{figure}
\begin{figure}[!htbp]
	\centering
	\includegraphics[scale = 0.5]{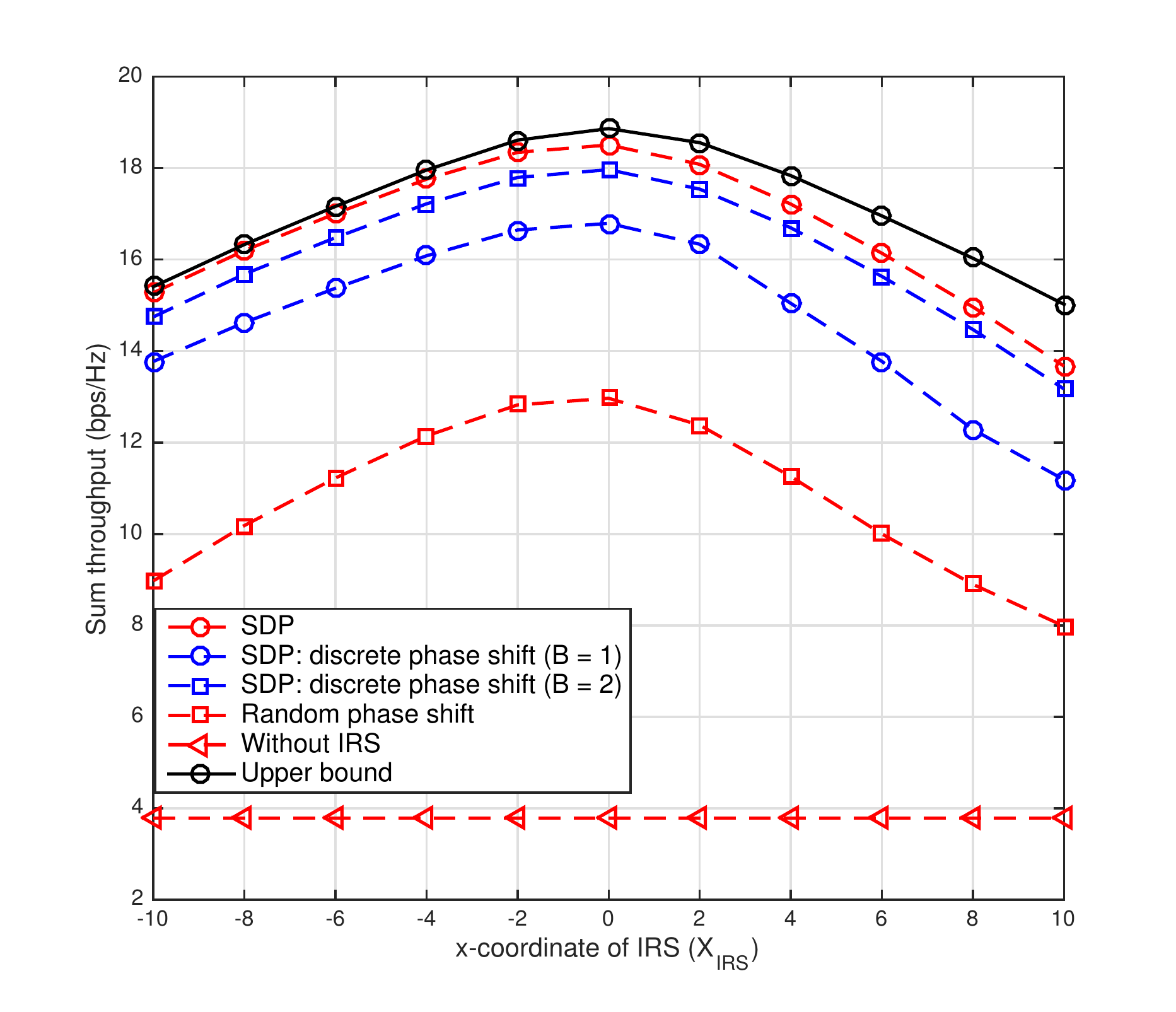}
	\caption{Sum throughput versus x-coordinate of IRS $ X_{\textrm{IRS}} $ for general case.}
	\label{fig:Rate_vs_X_IRS_general}
\end{figure}
Next, the impact of the IRS deployment on the sum throughput is illustrated for the special and general cases in Fig. \ref{fig:Rate_vs_X_IRS_special} and Fig. \ref{fig:Rate_vs_X_IRS_general}, respectively. It is seen from these results that the sum throughput first increases and then declines with x-coordinate of the IRS $ X_{\textrm{IRS}} $. This justifies the optimal deployment of the IRS to maximize the energy collection reception at the sensor nodes and the information reception at the AP, respectively. Also, the scheme with continuous phase shifts slightly outperforms its counterpart with discrete phase shifts, all of which have an obvious advantage over the schemes with random phase shifts and the one without IRS. In addition, a larger number of bits $ B $ yields a better performance, and the proposed low complexity scheme is closely matched with the SDP relaxed scheme.

\begin{figure}[!htbp]
	\centering
	\includegraphics[scale = 0.5]{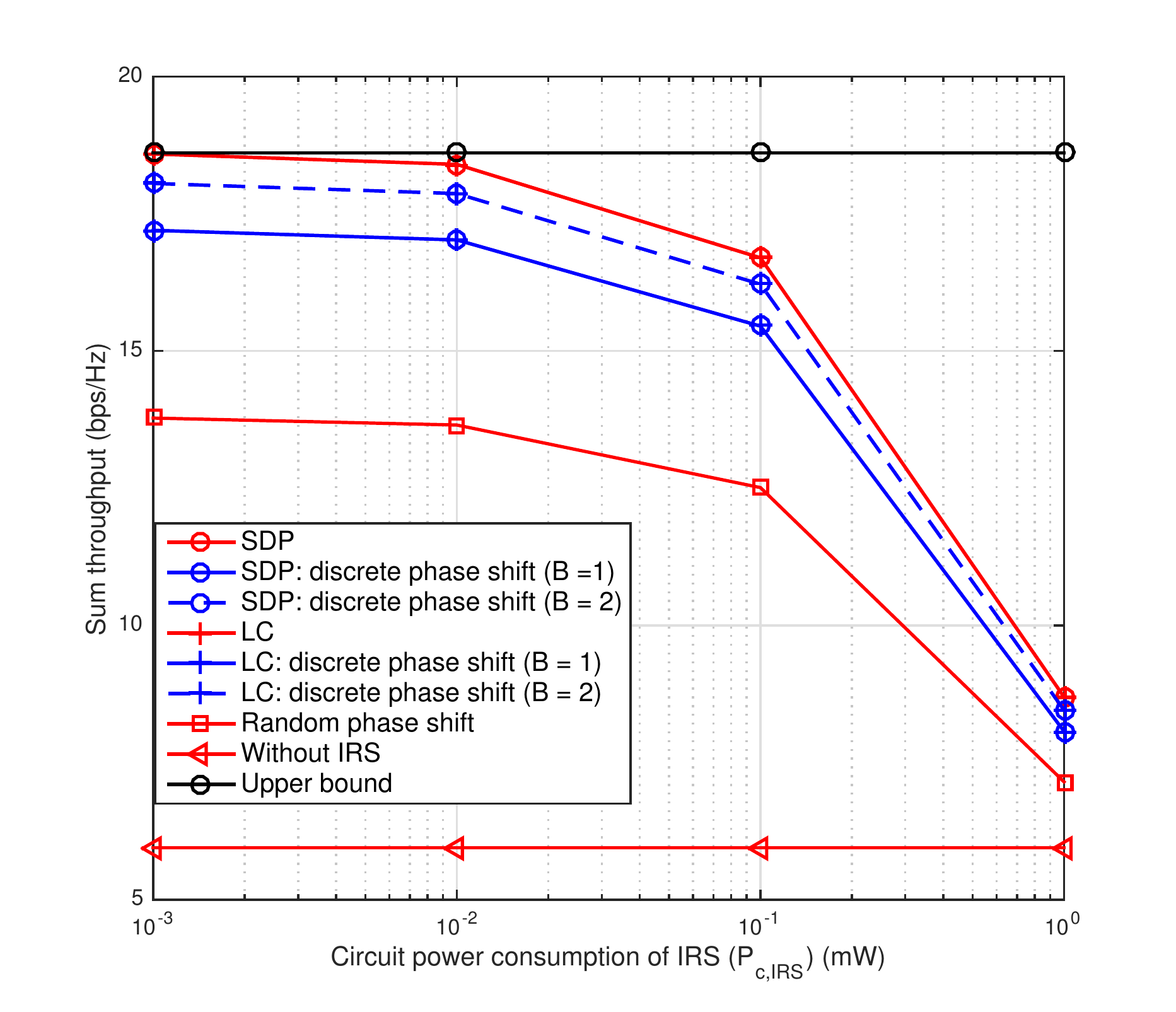}
	\caption{Sum throughput versus circuit power consumption of IRS $ P_{c,\textrm{IRS}} $ for special case.}
	\label{fig:Rate_vs_Pc_IRS_special}
\end{figure}
\begin{figure}[!htbp]
	\centering
	\includegraphics[scale = 0.5]{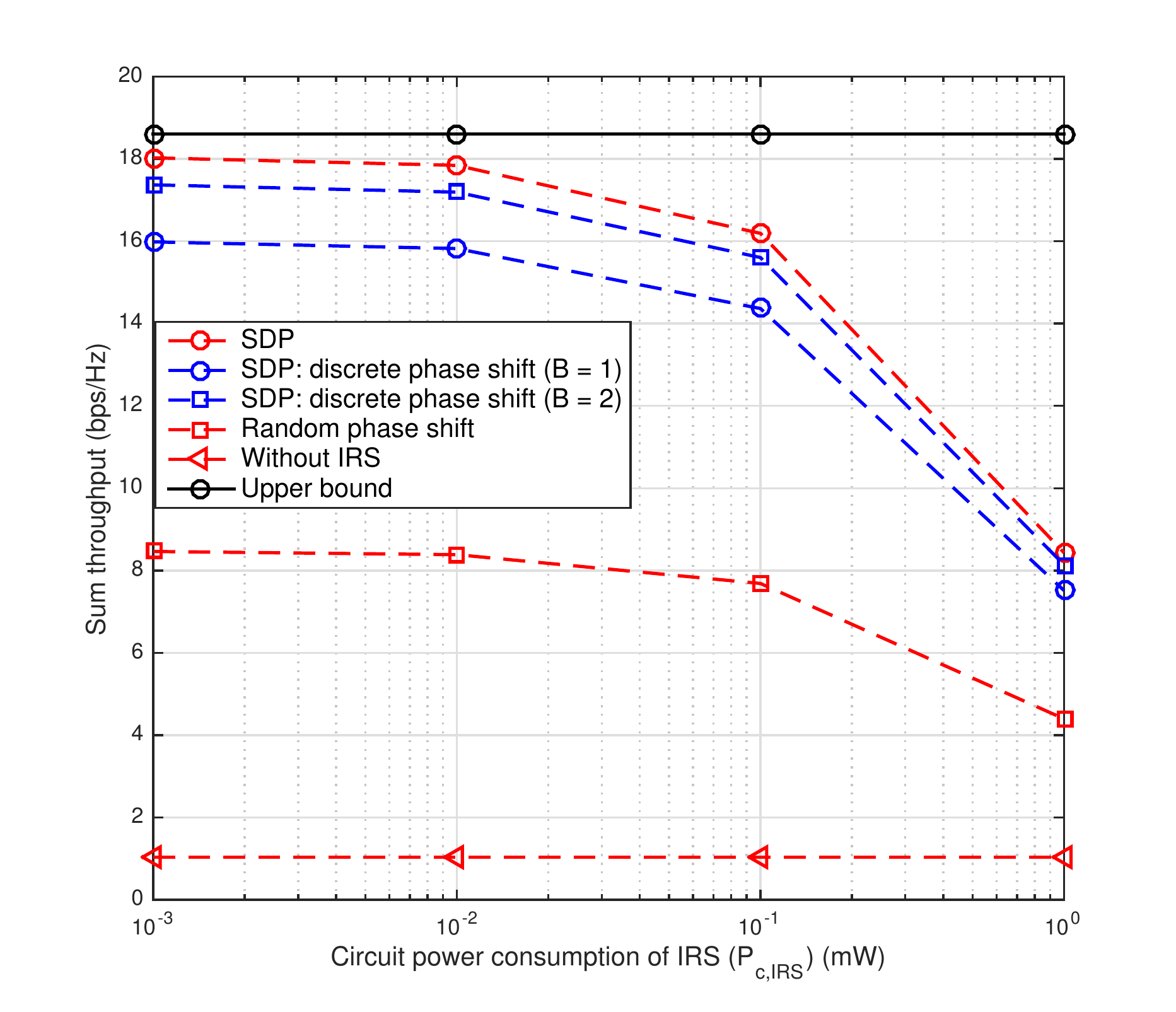}
	\caption{Sum throughput versus circuit power consumption of IRS $ P_{c,\textrm{IRS}} $ for general case.}
	\label{fig:Rate_vs_Pc_IRS_general}
\end{figure}
We evaluate the sum throughput versus the circuit power consumption of the IRS $ P_{c,\textrm{IRS}} $ for the special and general cases in Fig. \ref{fig:Rate_vs_Pc_IRS_special} and Fig. \ref{fig:Rate_vs_Pc_IRS_general}, respectively. As $ P_{c,\textrm{IRS}} $ increases, the sum throughput has an decreasing trend, which is more significant at the higher circuit power region. This unveils the fact that the IRS requires more energy to support its own circuit operation at the first sub-slot of the WET, thus cuts down its energy reflection at the second sub-slot of the WET. 

\begin{figure}[!htbp]
	\centering
	\includegraphics[scale = 0.5]{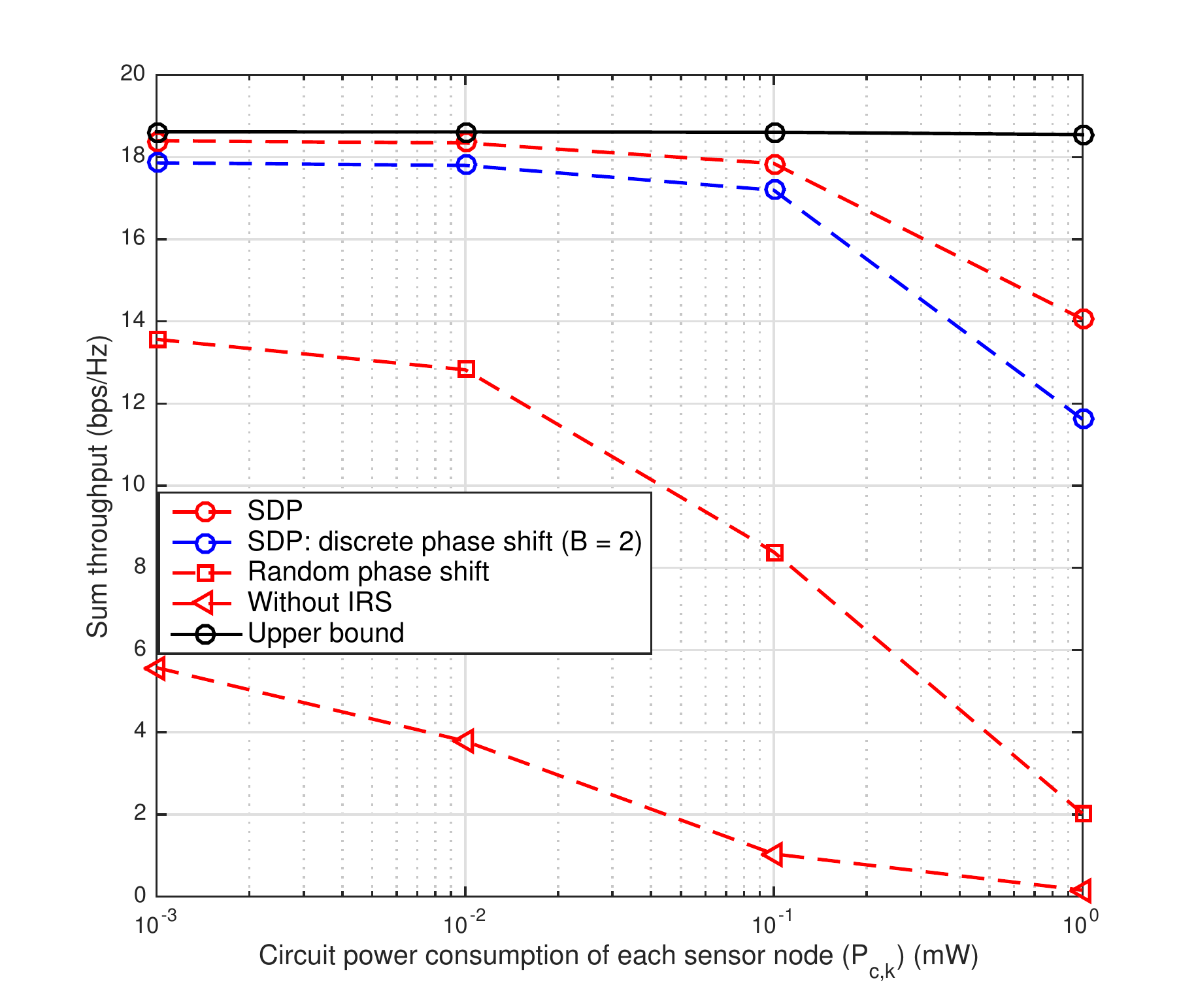}
	\caption{Sum throughput versus circuit power consumption of sensor nodes  $ P_{c,k} $ for general case.}
	\label{fig:Rate_vs_Pck_general}
\end{figure}
Moreover, we evaluate the impact of circuit power consumption of the sensor nodes $ P_{c,k} $ on sum throughput for the general case in Fig. \ref{fig:Rate_vs_Pck_general}.\footnote{We do not need to evaluate the sum throughput versus the circuit power consumption of the sensor nodes $ P_{c,k} $, since the special case is not affected by $ P_{c,k} $.}
From this result, we observe that the sum throughput decreases with $ P_{c,k} $, which becomes more evident in the higher circuit power region. This reveals the fact that the sensor nodes require more energy to support its own circuit operation so as to cut down its portion of energy for the WIT. 

\begin{figure}[!htbp]
	\centering
	\includegraphics[scale = 0.53]{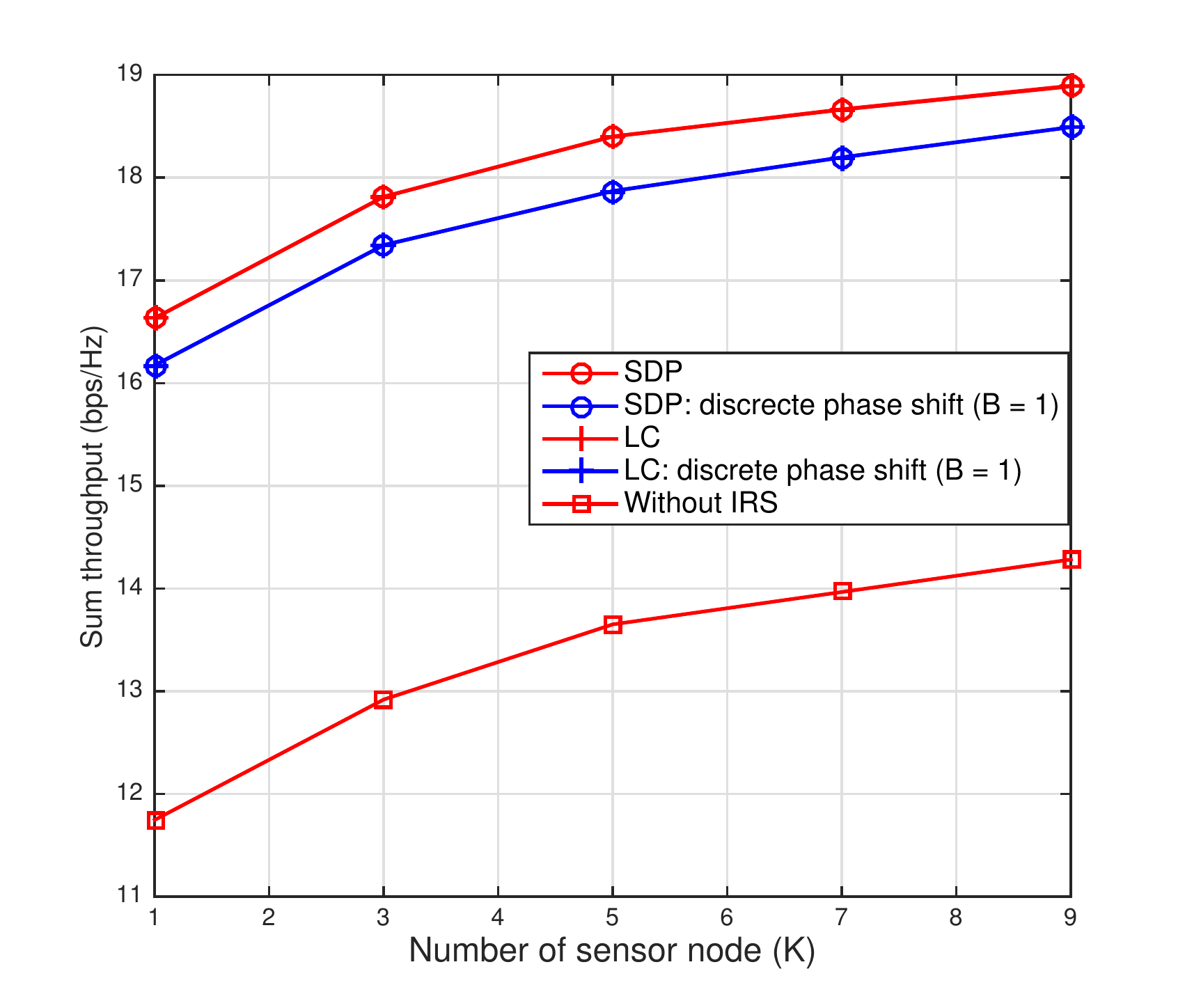}
	\caption{Sum throughput versus number of sensor nodes ($ K $).}
	\label{fig:Rate_vs_K_special}
\end{figure}
Finally, the impact of the number of sensor nodes $ K $ on the sum throughput performance is illustrated in Fig. \ref{fig:Rate_vs_K_special}, where the sum throughput has an increasing relation with the number of sensor nodes $ K $. It is also worth mentioning that, as $ K $ increases, more sensors are located farther away from the ES, the IRS, and the AP (refer to the system deployment in Fig. \ref{fig:System_deployment}), which degrades the energy and information reflection efficiencies, resulting in that the throughput gain becomes smaller. 

\section{Conclusion}\label{section:Conclusion}
This studied an IRS assisted WPSN, where the IRS aims to harvest energy from an ES to support its circuit operations and reflect energy and information signals to enhance WET and WIT performance.
We proposed a novel IRS-HTT-TS transmission policy to schedule the time slots during the WET phase. 
We are interested in maximizing the system sum throughput to design the discrete phase shifts of the WET and WIT stages, the power allocation of the sensors, and the transmission time scheduling. We first relaxed the original problem to the one with continuous phase shifts to deal with its NP-hard nature. To solve the relaxed non-convex problem, a two-step approach was proposed to decompose it into two sub-problems, which can be independently solved. The optimal closed-form phase shifts of the WIT stage were derived in closed-form. 
Furthermore, we specifically investigated two cases without/with the circuit power consumption of each sensor to optimally design the phase shift for the WET, the power allocation of each sensor as well as the transmission time scheduling. 
In addition, we considered the quantization of the continuous phase shifts to obtain the optimal discrete counterparts. Finally, the benefits of the proposed schemes were validated by numerical evaluations and shown to significantly enhance the sum throughput performance of the IRS assisted WPSN under study. 

The proposed approach has established an analytical framework for further investigation on the IRS aided WPSN, where the system model, transmission protocol and performance analysis can serve as a benchmark for the cases, such as, the network considered in this paper but instead with a non-linear energy harvesting model, where novel fractional energy harvesting models can be considered in the joint design of the phase shifts of the WET and WIT phases. Correspondingly, an extended system model with a multi-antenna ES or AP, where the active energy or active receiving and passive reflecting beamformers, as well as the transmission time slots can be alternately optimized. In addition, the impact of channel estimation errors, pilot overhead and hardware impairments on the throughput performance may be introduced by leveraging channel estimation algorithm, which is considered as one of our future work. Specifically, robust design will be studied in the IRS assisted WPSN to cope with the imperfect channel estimation, where the direct and cascaded channel estimation errors can be modelled as deterministic or statistical quantities.
Moreover, the IRS can be applied to mobile edge computing or wireless edge caching to improve the computational or wireless caching capabilities, which forges a novel paradigm for an integrated communication, sensing and computing network.
\begin{appendix}
\subsection{Proof of \emph{Theorem} \ref{lemma:EQphaseshifts}}\label{appendix:EQphaseshifts}
	According to \eqref{eq:Rk}, it is readily verified that $ R_{k} $ is a monotonically increasing function in the term $ \left | \mathbf{h}_{k} \bm{\Theta}_{k} \mathbf{h}_{r} + h_{d,k} \right |^{2} $. Thus, the optimal solution of $ \bm{\Theta}_{k} $ is obtained by
	solving problem \eqref{eq:Ori_problem_HTT_TS}, which is equivalent to solving the following $ K $ sub-problem
	\begin{align}\label{eq:Problem_WITphaseshifts}
		\max_{ \bm{\Theta}_{k} } &~ \left | \mathbf{h}_{k} \bm{\Theta}_{k} \mathbf{h}_{r} + h_{d,k} \right |^{2},~
		s.t. ~ \left | \exp \left( j \alpha_{k,n} \right) \right | = 1, \nonumber\\&~ \alpha_{k,n} \in [0, 2\pi), ~ \forall k \in [1,K],~\forall n \in [1,N_{R}].
	\end{align}	
	 Problem \eqref{eq:Problem_WITphaseshifts} only relies on $ \bm{\Theta}_{k} $ with $ \left | \exp \left( j \alpha_{k,n} \right) \right | = 1 $ and $ \alpha_{k,n} \in [0, 2\pi) $. 
	To solve \eqref{eq:Problem_WITphaseshifts}, we have
	\begin{align}\label{eq:Equality_WITphase}
		\left | \mathbf{h}_{k} \bm{\Theta}_{k} \mathbf{h}_{r} + h_{d,k} \right |
		= \left |  \bm{\theta}_{k} \mathbf{b}_{k} + h_{d,k} \right |,
	\end{align}
	where $ \mathbf{b}_{k} = \textrm{diag}\left( \mathbf{h}_{k} \right) \mathbf{h}_{r} $, $ \bm{\theta}_{k} = \left[ \theta_{k,1}, ... , \theta_{k,N_{R}} \right] = \left[ \exp\left( j \alpha_{k,1} \right) , ... , \exp \left( j \alpha_{k,N_{R}} \right)  \right] $, $ \left | \theta_{k,n} \right | = 1 $.
	To proceed, we apply the triangle inequality to the right hand side (RHS) of \eqref{eq:Equality_WITphase} to achieve its upper bound, which aims to obtain the optimal phase shifts of the WIT phase $ \bm{\theta}_{k} $ in problem \eqref{eq:Equality_WITphase} \cite{SZBi_IRS_WCL_2020}. As such, the upper bound of \eqref{eq:Equality_WITphase} is given as 
	\begin{align}\label{eq:UpperboundWITphase}
		\left |  \bm{\theta}_{k} \mathbf{b}_{k} \!+\! h_{d,k} \right | \! \leq \! \sum_{n = 1}^{N_{R}} \left | \theta_{k,n} \mathbf{b}_{k} [n] \right | \!+\! \left | h_{d,k} \right | \!=\! \sum_{n = 1}^{N_{R}} \left| \mathbf{b}_{k} [n] \right| \!+\! \left| h_{d,k} \right|,
	\end{align}
	where $ \mathbf{b}_{k} [n] $ is the $ n $-th element of $ \mathbf{b}_{k} $, and equality holds with $ \left | \theta_{k,n} \right | = 1 $ for $ n \in [1,N_{R}] $. According to \cite{SZBi_IRS_WCL_2020}, the upper bound in \eqref{eq:UpperboundWITphase} can be obtained via 
	\begin{align}\label{eq:Optimal_phaseWIT}
		\!\!\!\!\alpha_{k,n}^{*} \!=\! \arg \left( h_{d,k} \right) \!-\! \arg \left( \mathbf{b}_{k} [n] \right), \textrm{and}~ \theta_{k,n}^{*} \!=\! \exp \left( j \alpha_{k,n} \right).
	\end{align}
	Thus, the optimal phase shifts of the WIT is derived by $ \bm{\theta}_{k}^{*} $ in \eqref{eq:Optimal_phaseWIT}, and equivalently the optimal solution to \eqref{eq:Problem_WITphaseshifts} (i.e., $ \bm{\Theta}_{k}^{*} $) is obtained from $ \bm{\theta}_{k}^{*} $. 
	Thus, we complete the proof of \emph{Theorem} \ref{lemma:EQphaseshifts}.
	\subsection{Proof of \emph{Lemma} \ref{proposition:Alignment}} \label{appendix:Alignment}
	We exploit a few of mathematical manipulations to expand the term $ \left | \mathbf{h}_{k} \bm{\Theta}_{k}^{*} \mathbf{h}_{r} + h_{d,k} \right |^{2} $, 
	\begin{align}\label{eq:Expansion1}
		& \left | \mathbf{h}_{k} \bm{\Theta}_{k}^{*} \mathbf{h}_{r} + h_{d,k} \right |^{2}  = \left | \mathbf{h}_{k} \bm{\Theta}_{k}^{*} \mathbf{h}_{r} \right |^{2}  + \left |  h_{d,k} \right |^{2} \nonumber\\ &~~~\!+\! 2 \left | \mathbf{h}_{k} \bm{\Theta}_{k}^{*} \mathbf{h}_{r} \right |  \left |  h_{d,k} \right |  \textrm{cos} \left( \textrm{arg} (h_{d,k}) \!-\! \textrm{arg} ( \mathbf{h}_{k} \bm{\Theta}_{k}^{*} \mathbf{h}_{r} ) \right), 
	\end{align}
	From \eqref{eq:Expansion1}, the term $ \left | \mathbf{h}_{k} \bm{\Theta}_{k}^{*} \mathbf{h}_{r} + h_{d,k} \right |^{2} $ obtains its maximum value if 
	\begin{align}\label{eq:Alignment}
		\textrm{cos} \left( \textrm{arg} (h_{d,k}) - \textrm{arg} ( \mathbf{h}_{k} \bm{\Theta}_{k}^{*} \mathbf{h}_{r} ) \right) = 1.
	\end{align}
	\eqref{eq:Alignment} indicates that the phases of both direct and reflecting links between the $k$-th sensor node and the AP are identical, i.e., $ \textrm{arg} (h_{d,k}) = \textrm{arg} ( \mathbf{h}_{k} \bm{\Theta}_{k}^{*} \mathbf{h}_{r} ) $, which completes the proof of \emph{Lemma} \ref{proposition:Alignment}.
	\subsection{Proof of \emph{Lemma} \ref{lemma:Handle1} } \label{appendix:Handle1}
	The objective function \eqref{eq:Ori_problem_HTT_TS2_obj} is an increasing function with respect to $ P_{k} $ and $ \tau_{k},~\forall k \in [1,K] $. Thus, the constraints \eqref{eq:HTT_TS_specialsensorpower} and \eqref{eq:HTT_TS_IRSpower} are satisfied with equality when the optimal solutions to problem \eqref{eq:Ori_problem_HTT_TS2} are achieved. 
    From the constraint \eqref{eq:HTT_TS_IRSpower}, it is readily observed that its RHS is monotonically increasing with respect to $ t_{0,1} $. As such, the constraint \eqref{eq:HTT_TS_time} holds the equality with the optimal solution, otherwise it is always required to increase $ t_{0,1} $ such that $ t_{0,2} $ and the harvested energy of the IoT sensors can also be increased until the equality holds. Similarly,  the equality holds in the constraint \eqref{eq:HTT_TS_timeallocation0} with the optimal solution, otherwise it requires increasing $ \tau_{0} $, leading
	to the increase of $ t_{0,i} $ until equality holds.
	Thus, the following equalities hold with the optimal solution: 
	\begin{subequations}
		\begin{align}
			&   \tau_{k}^{*}  P_{k}^{*}   = t_{0,1}^{*} \eta P_{0} | g_{d,k} |^{2} + t_{0,2}^{*} \eta P_{0} | \mathbf{g}_{0} \bm{\Theta}_{0}^{*} \mathbf{g}_{r,k} + g_{d,k} |^{2}, \label{eq:EQ1} \\
			&	N_{R} P_{\textrm{c,IRS}} \left( t_{0,2}^{*} + \sum_{k=1}^{K} \tau_{k}^{*} \right)  = t_{0,1}^{*} \eta P_{0} \left\| \mathbf{g}_{0} \right\|^{2}, \label{eq:EQ2} \\ 
			&  \sum_{i = 1}^{2} t_{0,i}^{*}  = \tau_{0}^{*}, \label{eq:EQ3} \\ 
			& \sum_{k=0}^{K} \tau_{k}^{*}  = T, \label{eq:EQ4}
		\end{align}
	\end{subequations}
	From \eqref{eq:EQ1}, \eqref{eq:EQ3}, and \eqref{eq:EQ4}, the first part of \emph{Lemma} \eqref{lemma:Handle1} has been proved. We substitute \eqref{eq:Optimal_t020} into \eqref{eq:EQ2}, and then achieve the optimal closed-form $ t_{0,1}^{*} $ shown in \eqref{eq:Optimal_t01}, which completes the second part of \emph{Lemma} \eqref{lemma:Handle1}.
	\subsection{Proof of \emph{Theorem} \ref{theorem:TS_tauk}} \label{appendix:TS_tauk}
	To prove \emph{Theorem} \ref{theorem:TS_tauk}, we consider the Lagrange dual function of problem \eqref{eq:Ori_problem_HTT_TS_special2}, 
	\begin{align}
		\mathcal{L} ( t_{0,2}, \tau_{k}, \mu ) &~= \sum_{k=1}^{K} \tau_{k} \log \left( 1 + \frac{ \tilde{c}_{k}  \left( \bar{c}_{k} + t_{0,2} s_{k} \right) }{ \tau_{k}  } \right) \nonumber\\&~~~~- \mu \left( t_{0,1}^{*} + t_{0,2} + \sum_{k=1}^{K} \tau_{k} - T \right),
	\end{align}
	where $ \mu \geq 0 $ denotes the dual variable associate constraint \eqref{eq:Ori_problem_HTT_TS_special2_timeconstraint}. Additionally, the associated dual problem to \eqref{eq:Ori_problem_HTT_TS_special2} is given as 
	\begin{align}\label{eq:Dualproblem}
		\min_{ \{ t_{0,2}, \tau_{k} \} \in \mathcal{S} } \mathcal{L} ( t_{0,2}, \tau_{k}, \mu ),
	\end{align}
	where $ \mathcal{S} $ is the feasible set of any $ t_{0,2} $ and $ \tau_{k} $, $ \forall k \in [1,K] $, and has been shown
	in the constraints \eqref{eq:Ori_problem_HTT_TS_special2_timeconstraint} and \eqref{eq:Ori_problem_HTT_TS_special2_timeconstraint1}. As mentioned earlier, \eqref{eq:Ori_problem_HTT_TS_special2} can be relaxed into a convex problem that satisfies Slater's condition. Thus, the strong duality holds between \eqref{eq:Ori_problem_HTT_TS_special2} and \eqref{eq:Dualproblem} such that the optimal solution to
	\eqref{eq:Ori_problem_HTT_TS_special2} satisfies the KKT conditions, which is given by
	\begin{subequations}
		\begin{align}
			\mu^{*} \left( t_{0,1}^{*} + t_{0,2}^{*} + \sum_{k=1}^{K} \tau_{k}^{*} - T \right) = 0, \label{eq:KKT1} \\
			\frac{ \partial \mathcal{L} }{ \partial \tau_{k} } = 0. \label{eq:KKT2} 
		\end{align}
	\end{subequations}
	From \eqref{eq:KKT1}, we have $ \mu > 0 $ due to $ \sum_{i = 1}^{2} t_{0,i}^{*} + \sum_{k=1}^{K} \tau_{k}^{*} = T $ from \eqref{eq:Optimal_t020}. In addition, we exploit \eqref{eq:KKT2} to have
	\begin{align}\label{eq:KKT2_derivative}
		\log \left(\! 1 \!+\! \frac{ \tilde{c}_{k}  \left( \bar{c}_{k} \!+\! t_{0,2} s_{k} \right) }{ \tau_{k}  } \! \right)  \!-\! \frac{ \tilde{c}_{k}  \left( \bar{c}_{k} \!+\! t_{0,2} s_{k} \right) }{ \tau_{k} \!+\! \tilde{c}_{k}  \left( \bar{c}_{k} \!+\! t_{0,2} s_{k} \right) } \!=\! \mu.
	\end{align}
	The equality \eqref{eq:KKT2_derivative} is similar to the form $ f(x) = \log (1 + x) + \frac{x}{1 + x} $ which is a monotonically increasing function with respect to $ x $. Thus, in order to satisfy the above $ K $ equations in \eqref{eq:KKT2_derivative}, we have 
	\begin{align}
		\frac{ \tilde{c}_{1}  \left( \bar{c}_{1} + t_{0,2} s_{1} \right) }{ \tau_{1}  } = , ..., = \frac{ \tilde{c}_{K}  \left( \bar{c}_{K} + t_{0,2} s_{K} \right) }{ \tau_{K}  }
	\end{align}
	Denoting $
	\frac{ \tilde{c}_{k}  \left( \bar{c}_{k} + t_{0,2} s_{k} \right) }{ \tau_{k}  } = \frac{1}{r},
	$ we have
	\begin{align}\label{eq:tauk}
		\tau_{k} = r \tilde{c}_{k}  \left( \bar{c}_{k} + t_{0,2} s_{k} \right).
	\end{align}
	By substituting \eqref{eq:tauk} into \eqref{eq:Ori_problem_HTT_TS_special2_timeconstraint}, we obtain
	\begin{align}\label{eq:r}
		r = \frac{ T - t_{0,1}^{*} - t_{0,2} }{ \sum_{k=1}^{K} \tilde{c}_{k}  \left( \bar{c}_{k} + t_{0,2} s_{k} \right) }.
	\end{align}
	We substitute \eqref{eq:r} into \eqref{eq:tauk} to complete the proof of \emph{Theorem} \ref{theorem:TS_tauk}.
	\subsection{Proof of \emph{Theorem} \ref{theorem:Optimal_phase_WET}} \label{appendix:Optimal_phase_WET}
	To solve \eqref{eq:Subproblem_LC_min}, we approximate its objective function via the MM algorithm.
	\begin{proposition}\label{proposition:Surrogate_MM}\cite{Palomar_TSP_2016}
		The objective function \eqref{eq:Subproblem_LC_min_obj} is approximately written, for any given $ \bm{\theta^{(m)}} $ at the $ m $-th iteration and for any feasible $ \bm{\theta}_{0} $, as
		\begin{align}\label{eq:Surrogate_function_MM}
			f(\bm{\theta}_{0}) & =  \bm{\theta}_{0} \bm{\Phi} \bm{\theta}_{0}^{H} - 2 \Re \{ \bm{\theta}_{0} \bm{\gamma} \} + d \nonumber\\ 
			&	\leq \bm{\theta}_{0} \bm{\Upsilon} \bm{\theta}_{0}^{H} - 2\Re \left\{ \bm{\theta}_{0} \left[ (\bm{\Upsilon} - \bm{\Phi}) \tilde{\bm{\theta}}_{0}^{H} + \bm{\gamma} \right] \right\} \nonumber\\&~~~~+  \tilde{\bm{\theta}}_{0}  (\bm{\Upsilon} - \bm{\Phi}) \tilde{\bm{\theta}}_{0}^{H} +  d \nonumber\\
			& = \lambda_{ \max } ( \bm{\Phi} ) \| \bm{\theta}_{0} \|^{2} - 2 \mathcal{R} \left\{ \bm{\theta}_{0} \left[ \left( \lambda_{ \max } ( \bm{\Phi} ) \mathbf{I}_{N \times N} \!-\! \bm{\Phi} \right) \tilde{\bm{\theta}}_{0}^{H} 
			\right.\right. \nonumber\\ &~~~~\left. \left. + \bm{\gamma} \right] \right\} + \tilde{d}
			= g(\bm{\theta}|\bm{\theta}^{(m)}),
		\end{align}
		where $ \tilde{d} =  \tilde{\bm{\theta}}_{0} \left[ \lambda_{ \max } ( \bm{\Phi} ) \mathbf{I}_{N_{R} \times N_{R}} - \bm{\Phi} \right] \tilde{\bm{\theta}}_{0}^{H} + d $, and $ \bm{\Upsilon} = \lambda_{\max}( \bm{\Phi} ) \mathbf{I}_{N_{R} \times N_{R}} $.
	\end{proposition}
	\emph{Proposition} \ref{proposition:Surrogate_MM} constructs a surrogate function of \eqref{eq:Subproblem_LC_min_obj}, and it is easily verified that $ g (\bm{\theta}|\bm{\theta}^{(m)}) $ in \eqref{eq:Surrogate_function_MM} guarantees the conditions in \eqref{eq:MM_three_conditions}.
	\begin{align}\label{eq:Problem_reformulation_LC1}
		\min_{\bm{\theta}_{0}} &~ \lambda_{ \max } ( \bm{\Phi} ) \| \bm{\theta}_{0} \|^{2} - 2 \mathcal{R} \left\{ \bm{\theta}_{0} \tilde{\bm{\gamma}} \right\}  \nonumber\\
		s.t. &~ |\bm{\theta}_{0}(n)| = 1, ~\forall n \in [1,N_{R}].
	\end{align}
	It is clearly seen that $ \|\bm{\theta}_{0} \|^2 = N_{R} $ due to $| \bm{\theta}_{0}(n) | = 1 $.
	The term $ \mathcal{R} \left\{ \bm{\theta}_{0} \tilde{\bm{\gamma}} \right\} $ can be maximized when the phases of $ \bm{\theta}_{0}(n) $ and $ \tilde{\bm{\gamma}}(n) $ are identical. We complete the proof of \emph{Theorem} \ref{theorem:Optimal_phase_WET}.
	\subsection{Proof of \emph{Theorem} \ref{theorem:Optimal_t02}}\label{appendix:Optimal_t02}
	To prove \emph{Theorem} \ref{theorem:Optimal_t02}, we first rewrite the objective function \eqref{eq:Ori_problem_HTT_TS_special8_obj} with respect to $ t_{0,2} $ as 
	\begin{align}\label{eq:ft02}
		f(t_{0,2}) =  \left( \tilde{T} - t_{0,2} \right) \log \left(  1 + \frac{ c+ t_{0,2} s }{  \tilde{T} - t_{0,2}  } \right)
	\end{align}
	Considering the first-order derivative of \eqref{eq:ft02} and setting it to zero, we have the following equality
	\begin{align}\label{eq:LamberW01}
		x \log \left( x \right) - x = s^{*} - 1,
	\end{align}
	where $ x = 1 + \frac{ c+ t_{0,2} s  }{  \tilde{T} - t_{0,2}  } $. With some mathematical manipulations, \eqref{eq:LamberW01} can be expressed equivalently as
	\begin{align}\label{eq:LamberW1}
		\log \left( \frac{x}{ \exp(1) } \right)  \exp \left[ \log \left( \frac{ x }{ \exp(1) } \right) \right] = \frac{ s - 1 }{ \exp(1) },
	\end{align}
	Applying the relation $ z \exp (z) = y \Rightarrow z = \mathcal{W}(y) $ into \eqref{eq:LamberW1}, we complete the proof of \emph{Theorem} \ref{theorem:Optimal_t02}.
	\subsection{Proof of \emph{Theorem} \ref{theorem:Optimal_generalcase1}} \label{appendix:Optimal_generalcase1}
	To prove \emph{Theorem} \ref{theorem:Optimal_generalcase1}, the Lagrange dual function of \eqref{eq:HTT_TS_general_time} is applied, 
	\begin{align}
		\mathcal{L}\left( t_{0,2}, \tau_{k}, \mu_{1} \right) &~= \sum_{k=1}^{K} \tau_{k} \log \left( 1 + \frac{ a_{k}+ t_{0,2} b_{k} }{ \tau_{k} }  - d_{k}  \right) \nonumber\\&~~~~~ - \mu_{1} \left(  t_{0,1}^{*} + t_{0,2} + \sum_{k=1}^{K} \tau_{k} - T \right),
	\end{align}
	where $ \mu_{1} $ denotes the dual multiplier associated with constraint \eqref{eq:Timeconstraint}. Taking the first-order partial derivative of this dual function with respect to
	$ t_{0,2} $ and $ \tau_{k} $, respectively, we have 
	\begin{subequations}
		\begin{align}
			\frac{ \partial \mathcal{L} }{ \partial t_{0,2} } &= \sum_{k=1}^{K} \frac{ b_{k} }{ 1 + x_{k} - d_{k} } - \mu_{1}, \label{eq:Derivative_t02}\\
			\frac{\partial \mathcal{L}}{\partial \tau_{k}} &= \log \left( 1 + x_{k} - d_{k} \right) - \frac{ x_{k}  }{ 1 + x_{k} - d_{k} } - \mu_{1}, \label{eq:Derivative_tauk}
		\end{align}
	\end{subequations}
	where $ x_{k} = \frac{ a_{k} + t_{0,2}b_{k} }{ \tau_{k} },~\forall k \in [1,K] $. To obtain the optimal closed-form solution of $ t_{0,2} $ and $ \tau_{k},~\forall k \in [1,K] $, we let $ \frac{ \partial \mathcal{L} }{ \partial t_{0,2} } = 0 $ and $ \frac{\partial \mathcal{L}}{\partial \tau_{k}} = 0,~\forall k \in [1,K] $, respectively. Consequently, the optimal solution $ x_{k}^{*}  $ can be obtained via solving the following equation 
	\begin{align}
		&\!	\log \left( 1 \!+\! x_{k} \!-\! d_{k} \right) \!-\! \frac{ x_{k}  }{ 1 \!+\! x_{k}  \!-\! d_{k} } \!=\! \sum_{k=1}^{K} \frac{ b_{k} }{ 1 \!+\! x_{k}  \!-\! d_{k} },\forall k \! \in \! [1,K], \nonumber 
	\end{align}
	which can be efficiently obtained by numerical methods, e.g., bisection method. Then, the constraint \eqref{eq:Timeconstraint} holds with equality at the optimal solution, and we have
	\begin{align}
		t_{0,1}^{*} + t_{0,2} + \sum_{k=1}^{K}  \frac{ a_{k} + t_{0,2} b_{k} }{ x_{k}^{*} } = T. 
	\end{align}
	Hence, the optimal closed-form time, $ t_{0,2}^{*} $ and $ \tau_{k},~\forall k \in [1,K] $ can be derived as \eqref{eq:tauk11}. This completes the proof of \emph{Theorem} \ref{theorem:Optimal_generalcase1}. 
	\end{appendix}

\bibliographystyle{ieeetr}
\bibliography{my_references}

\end{document}